\documentclass[12pt]{article}

\usepackage[dvips]{graphicx}
\newcommand{\gsim}{\raisebox{-4pt}{$\,\stackrel{\textstyle
                                                         >}{\sim}\,$}}
\newcommand{\nn}{\nonumber}
\newcommand{\be}{\begin{equation}}
\newcommand{\ee}{\end{equation}}
\newcommand{\ba}{\begin{eqnarray}}
\newcommand{\ea}{\end{eqnarray}}
\newcommand{\req}[1]{(\ref{#1})}
\def\={\,=\,}
\newcommand{\ci}[1]{\cite{#1}}

\def\mev{~{\rm MeV}}
\def\gev{~{\rm GeV}}

\def\ale{\alpha_{\rm elm}}
\def\als{\alpha_{\rm s}}
\def\eps{\epsilon}

\def\xbj{x_{\rm Bj}}

\def\xb{\bar{x}}
\newcommand{\tw}{\textwidth}
\def\vk{{\bf k}_{\perp}}

\def\vbs{{\bf b}}
\def\vb0{{\bf b}_0}

\def\xbj{x_{\rm Bj}}
\newcommand{\sla}{\hspace*{-0.20cm}/}

\newcommand{\wf}{wave function}

\def\={\,=\,}

\begin{document} 
%%%%%%%%%%%%%%%%%%%%%%%%%%%%%%%%%%%%%%%%%%%%%%%%%%%%%%%%%%%%%%%%%%%%%%%%%%%%%%%%%%%%%%%%%%%%%
\thispagestyle{empty}
\begin{flushright}
WU B 09-06 \\
%hep-ph/yymmnnn\\
August, 30 2009\\[20mm]
\end{flushright}

\begin{center}
{\Large\bf An attempt to understand exclusive $\pi^+$ electroproduction} \\
\vskip 10mm

S.V.\ Goloskokov
\footnote{Email:  goloskkv@theor.jinr.ru}
\\[1em]
{\small {\it Bogoliubov Laboratory of Theoretical Physics, Joint Institute
for Nuclear Research,\\ Dubna 141980, Moscow region, Russia}}\\
\vskip 5mm

P.\ Kroll \footnote{Email:  kroll@physik.uni-wuppertal.de}
\\[1em]
{\small {\it Fachbereich Physik, Universit\"at Wuppertal, D-42097 Wuppertal,
Germany}}\\
and\\
{\small {\it Institut f\"ur Theoretische Physik, Universit\"at
    Regensburg, \\D-93040 Regensburg, Germany}}\\

\centerline{(revised version )}
%\date
\end{center}
\vskip 5mm 
\begin{abstract}
Hard exclusive $\pi^+$ electroproduction is investigated within the handbag
approach. The prominent role of the pion-pole contribution is demonstrated.
It is also shown that the experimental data require a twist-3 effect
which ensues from the helicity-flip generalized parton distribution $H_T$ 
and the twist-3 pion wave function. The results calculated from this handbag 
approach are compared in detail with the experimental data on cross sections 
and spin asymmetries measured with a polarized target. It is also commented 
on consequences of this approach for exclusive $\pi^0$ and vector-meson
electroproduction.
\end{abstract}

%%%%%%%%%%%%%%%%%%%%%%%%%%%%%%%%%%%%%%%%%%%%%%%%%%%%%%%%%%%%%%%%%%%%%%%%%%%%%%%
\section{Introduction}
\label{sec:intro}
%%%%%%%%%%%%%%%%%%%%%%%%%%%%%%%%%%%%%%%%%%%%%%%%%%%%%%%%%%%%%%%%%%%%%%%%%%%
With the advent of new, rather precise data on hard exclusive
processes in recent years the interest in the theoretical analysis of
such processes within QCD has strongly grown. The basis of such
analyses is the factorization of the process amplitudes into hard
subprocesses and soft hadronic matrix elements parameterized in terms
of generalized parton distributions (GPDs). An indispensable purpose
of such analyses is the scrutiny of this so-called handbag approach
within the experimentally accessible kinematical region. It is to be 
examined for instance whether factorization holds or whether there
is need for higher-twist and/or power corrections. Another task is to extract
as much as possible qualitative and quantitative information about the
GPDs. These functions embody the internal partonic structure of the 
hadrons. In addition to the longitudinal momentum distribution of the
partons which is already provided by the usual parton distributions,
the GPDs also incorporate the transverse distributions of the partons
inside the proton. A fully model independent extraction of the GPDs from
experiment is presumably impossible. Therefore one is likely dependent on
models or parameterizations of the GPDs constraint by theoretical concepts.
Frequently used is an integral representation of the GPDs where the integrand
represents a the double distribution \ci{mul94,rad98}. This way the skewness
dependence of the GPDs is generated. Other concepts have been exploited    
in recent analyses of deeply virtual Compton scattering: In \ci{mueller09}
the GPDs are constraint by analyticity and in \ci{polyakov09} a dual
parameterization is employed. 

In our recent analyses of vector-meson electroproduction \ci{GK1,GK2,GK3} the 
double distribution method has been used. The rather small ratio of the 
longitudinal over transversal cross sections indicates the important role of 
transversely polarized virtual photons in the experimentally accessible range 
of photon virtualities. The transversely polarized photons manifestly lead to 
higher-twist and/or power corrections which are asymptotically suppressed by 
the inverse of the photon virtuality, $1/Q$, in the amplitudes. 
Here in this work we attempt an analysis of hard exclusive electroproduction
of charged pions. The description of this process within the handbag approach
presents a major challenge.  Besides the differential cross sections measured
by the $F_\pi-2$ collaboration \ci{horn06} and HERMES \ci{HERMES07} which
exhibit a pronounced forward spike, there are also data on spin asymmetries
measured with a polarized target \ci{Hristova,hermes02}. Spin dependent
observables typically probe subtle details of the amplitudes as for
instance relative phases. The pertinent GPDs occurring in $\pi^+$
electroproduction are the poorly known $\widetilde{H}$ and $\widetilde{E}$. 
This is to be contrasted with vector-meson electroproduction where mainly the 
GPDs $H$ and $E$ are probed. Electroproduction of pions has been investigated 
several times \ci{man98,man99,frankfurt99,VGG} to leading-twist accuracy. It 
turns out that this naive although theoretically clean approach, fails by
order of magnitude with the experimental cross section. The main reason for 
this disaster is the use of only the perturbative contribution to the pion
form factor which is mandatory to leading-twist accuracy.  
 
In the present analysis of $\pi^+$ electroproduction we are going to vary from
the leading-twist approximation in several aspects. First we will make use
of the modified perturbative approach \ci{botts89} for the calculation
of the subprocess amplitudes as we did for vector-meson electroproduction. 
Secondly we will use the full electromagnetic form factor of the pion instead 
of only its much smaller perturbative contribution. Thirdly, we will allow for 
contributions from the GPD $\widetilde{E}$ and lastly, we will take into 
account a twist-3 effect which involves the helicity-flip GPD $H_T$ \ci{diehl01}.  
As we are going to demonstrate below the data on the spin asymmetry obtained
with a transversely polarized target \ci{Hristova} demand this
contribution. With respect to the fact that we need three (almost)
unknown GPDs for exclusive $\pi^+$ electroproduction we consider our
analysis as a rather qualitative study, not all aspects of the data
will be accommodated well. Yet we think of our work as the next step towards
a comprehensive analysis of hard exclusive meson electroproduction.

The plan of the paper is the following: In the next section we will
briefly sketch the description of the leading longitudinal amplitude
within the handbag approach. In Sect.\ \ref{sec:pole} we will present the
pion-pole contribution. In Sect.\ \ref{sec:transversal} we will demonstrate
the need for contributions from transversely polarized  virtual
photons in $\pi^+$ production and thereafter describe the twist-3
mechanism (Sect.\ \ref{sec:twist-3}). In Sect.\ \ref{sec:GPD} we will
present the model GPDs and the pion wave functions. Our results are
compared to experiment in Sect.\ \ref{sec:fits}. Sects.\
\ref{sec:pi0} and \ref{sec:vector} are devoted to a brief discussion
of implications of our approach for exclusive $\pi^0$ and vector-meson
electroproduction, respectively. A remarks on a symmetry property of
the helicity amplitudes are presented in Sect.\ \ref{sec:symmetry}. We
close the paper with a summary (Sect.\ \ref{sec:summary}). In an
Appendix we compile the relations between various observables of $\pi$
electoproduction and helicity amplitudes.      

%%%%%%%%%%%%%%%%%%%%%%%%%%%%%%%%%%%%%%%%%%%%%%%%%%%%%%%%%%%%%%%%%%%%%%%%%%%%%%%%
\section{The leading amplitudes}
\label{sec:leading}
%%%%%%%%%%%%%%%%%%%%%%%%%%%%%%%%%%%%%%%%%%%%%%%%%%%%%%%%%%%%%%%%%%%%%%%%%%%%%%%%
In the GPD-based handbag factorization scheme it can be shown \ci{col96} that
for large photon virtualities but small invariant momentum transfer, $t$, from 
the incoming to the outgoing baryon, the amplitudes for longitudinally
polarized photons ${\cal M}_{0\nu^\prime,0\nu}$ dominate the process 
$\gamma^* p\to \pi^+ n$. Here, $\nu$ and $\nu^\prime$ label the helicities of
the proton and neutron, respectively. The amplitudes for transversely
polarized photons are suppressed by $1/Q$ against the longitudinal ones.

Within the handbag approach the amplitudes for pion electroproduction through
longitudinally polarized photons read
\ba
{\cal M}^{\pi^+}_{0+,0+} &=& \sqrt{1-\xi^2}\, \frac{e_0}{Q}
                             \,\Big[\langle \widetilde{H}^{(3)}\rangle
-\frac{\xi^2}{1-\xi^2}\langle \widetilde{E}^{(3)}\rangle 
  - \frac{2\xi mQ^2}{1-\xi^2}\frac{\rho_\pi}{t-m_\pi^2}\Big]\,,\nn\\
{\cal M}^{\pi^+}_{0-,0+} &=& \frac{e_0}{Q}\,\frac{\sqrt{-t^\prime}}{2m}\,\Big[ \xi 
\langle \widetilde{E}^{(3)}\rangle + 2mQ^2\frac{\rho_\pi}{t-m_\pi^2}\Big]\,.
\label{eq:L-amplitudes}
\ea 
Here, the usual abbreviation $t^\prime=t-t_0$ is employed where 
$t_0=-4m^2\xi^2/(1-\xi^2)$ is the minimal value of $t$ corresponding to
forward scattering. The mass of the nucleon is denoted by $m$ and the skewness
parameter, $\xi$, is related to Bjorken-$x$ by 
\be
\xi\=\frac{\xbj}{2-\xbj}\,.
\ee
Helicity flips at the baryon vertex are taken into account since they are only
suppressed by $\sqrt{-t^\prime}/m$. In contrast to this, effects of order 
$\sqrt{-t^\prime}/Q$ are neglected.

\begin{figure}[t]
\begin{center}
\includegraphics[width=0.60\tw,bb=143 514 600 650,clip=true]{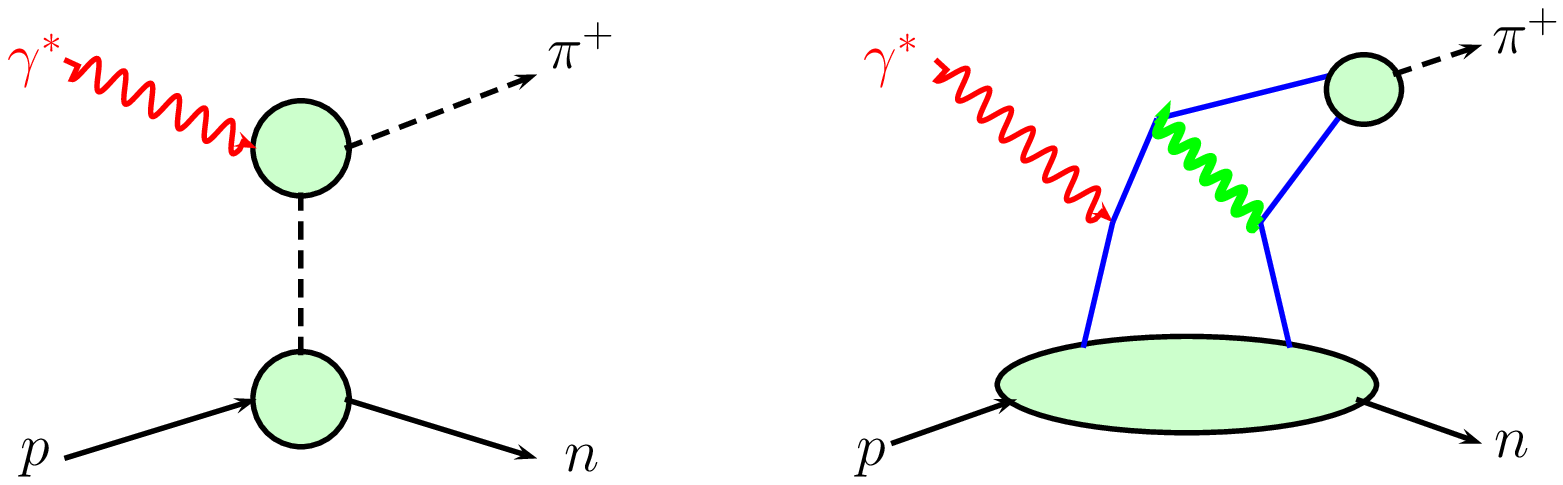}
\caption{\label{fig:graphs} The pion pole (left) and the handbag (right) contribution to
  electroproduction of positively charged pions.}
\end{center}
\end{figure}

The pion pole contribution, see Fig.\ \ref{fig:graphs}, with the residue
$\rho_\pi$ and pion mass $m_\pi$, will be discussed in the next section in
some detail. In constrast to other work \ci{man98,man99,frankfurt99,VGG}  
but in view of the fact that we take into account the full pion form factor 
and not just its perturbative contribution, we write the pion pole
contributions and those from  $\widetilde{E}^{(3)}$ separately, i.e.\ our  
$\widetilde{E}^{(3)}$ only represents the non-pole contribution of the full GPD.
For large $Q^2$, $\rho_\pi\propto 1/Q^2$ so that all terms in 
\req{eq:L-amplitudes} possess the same scaling behavior. The item 
$\langle \widetilde{F}\rangle$ denotes a convolution of a GPD 
$\widetilde{F} (= \widetilde{H}, \widetilde{E})$ and an appropriate subprocess
amplitude to be calculated from Feynman graphs of which a typical lowest 
order example is shown in Fig.\ \ref{fig:graphs},  
\be
\langle \widetilde{F}^{(3)} \rangle \= \sum_\lambda\int_{-1}^1 d\xb 
   {\cal H}_{0\lambda,0\lambda}(\xb,\xi,Q^2,t=0) \widetilde{F}^{(3)}(\xb,\xi,t)\,.
\label{eq:convolution}
\ee 
The label $\lambda$ refers to the unobserved helicities of the partons
participating in the subprocess. For $\pi^+$ production the $p\to n$
transition GPDs are required which are given by the isovector combination
of proton GPDs \ci{man99,frankfurt99}  
\be  
\widetilde{F}^{(3)} \= \widetilde{F}^u - \widetilde{F}^d\,.
\label{isovector}
\ee
If the sea-quark GPDs are flavor symmetric only the differences of the valence
quark GPDs contribute to $\pi^+$ electroproduction.

As in our previous work on hard electroproduction of vector mesons
\ci{GK1,GK2,GK3} we calculate the $\gamma^* q\to \pi q$ subprocess amplitudes 
within the modified perturbative approach \ci{botts89} in which quark
transverse degrees of freedom as well as Sudakov suppressions are taken into
account. Since the re-summation of the logarithms involved in the Sudakov
factor can only be performed in the impact parameter space efficiently \ci{botts89}
we quote the subprocess amplitudes in that space
\ba
{\cal H}_{0\lambda,0\lambda} &=& \int d\tau d^2b\, 
         \hat{\Psi}_{\pi}(\tau,-\vbs)\, 
      \hat{\cal F}^{(3)}_{0\lambda,0\lambda}(\xb,\xi,\tau,Q^2,\vbs)\, \nn\\ 
      && \times   \als(\mu_R)\,{\rm exp}{[-S(\tau,\vbs,Q^2)]}\,.
\label{mod-amp}
\ea
For the Sudakov factor $S$, the choice of the renormalization ($\mu_R$) and
factorization scales as well as the hard scattering kernels ${\cal F}$ or
their respective Fourier transforms $\hat{\cal F}$, we refer to Ref.\ \ci{GK3}. 
The last item in \req{mod-amp} to be explained is $\hat{\Psi}_{\pi}(\tau,-\vbs)$ 
which represents the Fourier transform of the momentum-space light-cone \wf{}
(LCWF) for the pion ($\tau$ is the momentum fraction of the quark that enters the
meson, defined with respect to the meson momentum). It is to be emphasized
that parton transverse momenta are only taken into account in the subprocess
while the partons entering the subprocess are viewed as being emitted and
re-absorbed by the nucleon collinearly. This approximation is justified to
some extent by the fact that the GPDs describe the full proton whereas the
meson is only generated through its compact valence Fock state. For a detailed
discussion of the modified perturbative approach and its application to hard 
meson electroproduction we refer to \ci{GK1,GK2,GK3}.
  
For the ease of comparison with other work we quote the momentum-space 
subprocess amplitude in collinear approximation where the pion \wf{} reduces 
to its associated leading-twist distribution amplitude:
\be
\sum_\lambda\,{\cal H}_{0\lambda,0\lambda}\= \frac{C_F}{N_c}4\pi\als(\mu_R)
     f_\pi \langle 1/\tau\rangle_\pi \Big[\frac{e_u}{\xb-\xi+\imath
       \varepsilon} +\frac{e_d}{\xb+\xi-\imath \varepsilon}\Big]\,,
\ee
where $f_\pi(=131\,\mev)$ denotes the pion decay constant and $\langle
1/\tau\rangle_\pi$ the $1/\tau$ moment of the pion's distribution
amplitude. The color factor $C_F$ is given by $(N_c^2-1)/(2N_c)$ where $N_c$
the number of colors. Finally, $e_a$ is the charge of quarks of flavor $a$ in 
units of the positron charge $e_0$. The running coupling $\als$ is evaluated
from the one-loop expression with $\Lambda_{\rm QCD}=240\,\mev$. 

%%%%%%%%%%%%%%%%%%%%%%%%%%%%%%%%%%%%%%%%%%%%%%%%%%%%%%%%%%%%%%%%%%%%%%%%%%%%%%%%%%
\section{The pion-pole contribution}
\label{sec:pole}
%%%%%%%%%%%%%%%%%%%%%%%%%%%%%%%%%%%%%%%%%%%%%%%%%%%%%%%%%%%%%%%%%%%%%%%%%%%%%%%%%%
The pion exchange graph shown in Fig.\ \ref{fig:graphs}, leads to the
following contribution to the helicity amplitudes of the process $\gamma^*
p\to \pi^+n$
\be
{\cal M}^{{\rm pole}}_{0\nu^\prime,\mu\nu} \= e_0\frac{\rho_\pi}{t-m^2_\pi}\,
 (2q^\prime-q)\cdot\epsilon(\mu)\, \bar{u}(p^\prime,\nu^\prime)\gamma_5 u(p,\nu)\,,
\label{eq:feynman}
\ee  
where $p, p^\prime$, $q^\prime$ and $q$ are the momenta of the proton, neutron,
pion and photon, respectively. The polarization vector of the virtual photon
is denoted by $\epsilon$ and its helicity by $\mu$. The residue of the pole is 
given by
\be
\rho_\pi\= \sqrt{2}g_{\pi NN} F_\pi(Q^2) F_{\pi NN}(t^\prime)\,.
\ee
The coupling of the pion to the nucleon is given by the familiar pion-nucleon
coupling constant, $g_{\pi NN}$, for which we take the value $13.4$. 
The structure of the pion and the nucleon is taken into account
by form factors, the electromagnetic one for the pion, $F_\pi(Q^2)$, 
whereby the small virtuality of the exchanged pion is as usual
ignored, and $F_{\pi NN}(t)$ for the $\pi$-nucleon vertex. 

Working out the spinor expression \req{eq:feynman}, one obtains for the
helicity amplitudes at small $-t$ and large $Q^2$\\
\ba
{\cal M}^{\rm pole}_{0+,0+} &=& - e_0 \frac{2m\xi Q}{\sqrt{1-\xi^2}}\, 
                           \frac{\rho_\pi}{t-m^2_\pi}\,,\nn\\
{\cal M}^{\rm pole}_{0-,0+} &=& + e_0 Q \sqrt{-t^\prime}\, \frac{\rho_\pi}{t-m^2_\pi}\,,\nn\\
{\cal M}^{\rm pole}_{0+,\pm +} &=& \pm 2\sqrt{2}e_0\xi m \sqrt{-t^\prime}\,
                             \frac{\rho_\pi}{t-m^2_\pi}\,,\nn\\
{\cal M}^{\rm pole}_{0-,\pm +} &=&
\pm \sqrt{2}e_0 t^\prime \sqrt{1-\xi^2}\,\frac{\rho_\pi}{t-m^2_\pi}\,. 
\label{eq:pion-pole-amplitude}
\ea
Terms suppressed by $1/Q^2$ against the longitudinal amplitudes, in particular 
terms of order $t^\prime/Q^2$, are neglected in order to be compatible with
the approximations used for the GPD contributions.

As one may see from the relations \req{eq:pion-pole-amplitude} the pion-pole
contributions to the amplitudes for transversely polarized photons vanish for
forward scattering. One would therefore expect a forward dip in the cross section
for photoproduction of pions or other pion-exchange dominated reactions like 
$p\bar{p}\to n\bar{n}$ or proton-neutron charge exchange. However, this
expectation is in sharp contrast to the behavior of the experimental cross
sections which rather exhibit pronounced forward spikes with widths of order 
of the pion mass squared. Hence, one is compelled to conclude that there is
another contribution that conspires with the pion contribution in such a way 
that a non-vanishing forward cross section is generated \ci{phillips}. A popular 
dynamical realization of such a conspirator is the so-called poor man's
absorption model \ci{williams} in which the $L=0$ partial wave of this
amplitude is absorbed completely. Effectively this prescription consists of 
replacing the factor $t^\prime$ in the helicity non-flip amplitude 
${\cal M}^{\rm pole}_{0-,++}$ by $m^2_\pi$. It is important to realize in this 
context that this amplitude has no net helicity flip. The factor $t^\prime$ 
occurring in it (cf.\ \req{eq:pion-pole-amplitude}) is a dynamical effect and 
not forced by angular momentum conservation~\footnote{
For $t^\prime \to 0$ a helicity amplitude vanishes (at least) as
${\cal M}_{\mu^\prime \nu^\prime,\mu\nu}\propto 
\sqrt{-t^\prime}^{\,|\mu-\nu-\mu^\prime+\nu^\prime|}$ 
as a consequence of angular momentum conservation.}.
The mentioned  replacement can be viewed as the net effect of adding a smooth
background to the pole term \ci{storrow}
\ba
{\cal M}^{\rm pole}_{0-,++} &\Longrightarrow &
{\cal M}^{\rm pole}_{0-,++} 
        + \sqrt{2} e_0 \rho_\pi \sqrt{\frac{1+\xi}{1-\xi}} \nn\\
    & \simeq &\sqrt{2} e_0
 (t_0-m^2_\pi)\,\sqrt{\frac{1+\xi}{1-\xi}} \frac{\rho_\pi}{t-m^2_\pi}\,.
\label{eq:pol-cut} 
\ea
This version of the amplitude ${\cal  M}^{\rm pole}_{0-,++}$ is a generalization 
of the familiar result for photoproduction of pions. Electromagnetic gauge 
invariance provides a further
argument for the version \req{eq:pol-cut}. Treating the pion and the nucleon
as point-like particles (or with common form factors) and adding the
contributions from nucleon exchange to the amplitudes \req{eq:pion-pole-amplitude}
\be
{\cal M}^N \= \frac{\sqrt{2} e_0g_{\pi NN}}{W^2-m^2}\, 
         \bar{u}(p^\prime,\nu^\prime)\gamma_5 (p\sla + q\sla +m)\, \eps\sla(\mu)
	 u(p,\nu)\,,
\ee 
(where $W$ the c.m.s. energy of the
process $\gamma^*p\to\pi^+ n$), one obtains a gauge invariant expression. At  
high energies however since $W^2-m^2\simeq Q^2(1-\xi)/(2\xi)$, all the
contributions from nucleon exchange are suppressed as compared to the amplitudes 
\req{eq:pion-pole-amplitude} with the exception of ${\cal  M}^N_{0-,++}$ which
has the form of the background term in \req{eq:pol-cut}. For a detailed
discussion of gauge invariance and Reggeization see \ci{leader71}. 
We are going to use \req{eq:pion-pole-amplitude} in our analysis with the
amplitude  ${\cal  M}_{0-,++}$ modified according to \req{eq:pol-cut}. 

It has been shown, for instance in Refs.\ \ci{VGG,penttinen}, that the pion pole
contribution can be viewed as part of the GPD $\widetilde{E}$: 
\be
 \widetilde{E}^{u}_{\rm pole}\=-\widetilde{E}^{d}_{\rm pole} \= \Theta(|\xb|\leq \xi)\, 
\frac{F_P(t)}{4\xi}\Phi_\pi((\xb+\xi)/(2\xi))\,,
\label{pole}
\ee
where $F_P$ is the pseudoscalar form factor of the nucleon being related
to $\widetilde{E}$ by the sum rule 
\be
\int_{-1}^1 d\xb \widetilde{E}^{(3)}(\xb,\xi,t) \= F_P(t)\,.
\ee
With the help of PCAC and the Goldberger-Treiman relation the pseudoscalar
form factor can be written as
\be
F_P(t)\= - m f_\pi \frac{2\sqrt{2}g_{\pi NN} F_{\pi NN}(t^\prime)}{t-m^2_\pi}\,.
\ee
Working out the convolution \req{eq:convolution} for 
$\widetilde{E}^{(3)}_{\rm pole}$ one exactly obtains the pion pole contribution
as given in \req{eq:L-amplitudes} with, however, the perturbative result for
the pion form factor obtained within the modified perturbative approach
\ci{jakob} or within its collinear approximation if one works with it. The
perturbative result underestimates the experimental value of the form factor
by about a factor of three for $Q^2$ in the range of $3-5\,\gev^2$. 

The measurement of the pion form factor bases on the fact that, at least at
large $\xi$ and small $-t$ , the longitudinal cross section for $\pi^+$ 
electroproduction, the process we are analyzing in this work, is dominated by 
the pion-pole contribution with some corrections from other sources (see
below). For instance, the Jefferson Lab $F_\pi-2$ collaboration \ci{horn06}  
has recently measured the longitudinal cross section and analyzed these data 
with a Regge parameterization \ci{vanderhaeghen} leaving $F_\pi(Q^2)$ as a 
free parameter. The fit to their data provides
\be
F_\pi(Q^2) \simeq \Big[1+Q^2/0.50\,\gev^2 \Big]^{-1}\,.
\label{pi-FF}
\ee
Obviously, the use of the perturbative result only is in conflict with the very
idea of measuring the pion form factor. We therefore refrain from using it in 
contrast to previous work, e.g.\ \ci{VGG}, and employ instead the experimental
value \req{pi-FF} of the pion's electromagnetic form factor in the evaluation
of the observables for $\pi^+$ electroproduction.

For the form factor of the pion-nucleon vertex we use the parameterization
\be
F_{\pi NN}(t^\prime)\=\big(\Lambda_N^2-m_\pi^2)/(\Lambda_N^2-t^\prime\big)\,,
\label{piN-FF}
\ee
where $\Lambda_N$ is considered as an adjustable parameter. 

Alternatively, one may use a reggeized version of pion exchange which is obtained
from \req{eq:pion-pole-amplitude} by the replacement
\be
\frac1{t-m_\pi^2} \Longrightarrow \frac{\pi\alpha'_\pi}{2}\,(\alpha_\pi(t)+1)\,
      \frac{1+\exp[-\imath\pi\alpha_\pi(t)]}{\sin{\pi\alpha_\pi(t)}}
          \left(\frac{W}{W_0}\right)^{2 \alpha_\pi(t)}\,.
\ee
For the pion trajectory one may take the usual linear form
\be
\alpha_\pi(t)\= \alpha'_\pi (t-m_\pi^2)\,,
\ee
with the slope $\alpha_\pi'=0.72\,\gev^{-1}$ as is fixed by the masses and the
spins of the $\pi$ and the $\pi_2(1670)$ mesons. The value of the slope is
somewhat smaller than the standard Regge slope of about $0.9\,\gev^{-1}$ owing
to  Goldstone boson nature of the pion. Without this symmetry effect the pion
mass would be roughly the same as the $\rho$ mass leading to the standard
slope. The scale $W_0$ may be chosen to be $1\,\gev$. The pion-nucleon form 
factor can be ignored in the reggeized version. Since, at large $Q^2$,
only data at practically constant value of $W$ are available \ci{HERMES07} 
both variants of pion exchange lead to fits of more or less the same quality. 
The results presented below are obtained with the non-reggeized version.  

%%%%%%%%%%%%%%%%%%%%%%%%%%%%%%%%%%%%%%%%%%%%%%%%%%%%%%%%%%%%%%%%%%%%%%%%%%%%%%%%%%
\section{Transversely polarized photons} %or The $\gamma_Tp\to \pi^+n$ amplitude}
\label{sec:transversal}
%%%%%%%%%%%%%%%%%%%%%%%%%%%%%%%%%%%%%%%%%%%%%%%%%%%%%%%%%%%%%%%%%%%%%%%%%%%%%%%%%%%
Here, in this section we want to argue that there is clear evidence for
contributions from transversely polarized photons in the existing data on hard
$\pi^+$ electroproduction.  

An information on contributions from such photons comes from the recent
Jefferson Lab measurement \ci{horn06} of the separated cross sections at,
however, the low energy of $\simeq 2.2\,\gev$. The transverse cross section is
sizeable; for  $-t\gsim 0.2\,\gev^2$ it is even larger than the longitudinal one. 
The importance of the pion pole is demonstrated in Tab.\
\ref{tab:2} where its contribution is confronted to the 
to the Jefferson Lab data \ci{horn06}. The pion-pole contribution is evaluated 
from \req{eq:pion-pole-amplitude} and \req{eq:pol-cut} with the help of
\req{pi-FF} and \req{piN-FF} with $\Lambda_N=0.51\,\gev$. We use this
value of  $\Lambda_N$ throughout this work. One observes that
the longitudinal cross section is indeed dominated by the pole contribution,
it even overshoots the data somewhat~\footnote{
The dominance of the pion contribution to the longitudinal cross section has
been also found in a recent Regge analysis of $\pi^+$ electroproduction \ci{mosel}.}. 
On the other hand, the pole contribution to the transverse cross section is
much smaller than experiment. There is clear evidence for other contributions
from transversely polarized photons.
\begin{table*}[t]
\renewcommand{\arraystretch}{1.4} 
\begin{center}
\begin{tabular}{| c | c | c || c | c | c |c|}
\hline     
 $W$  & $Q^2$ & $-t^\prime$  & $d\sigma_L/dt$ & pole  & $d\sigma_T/dt$ & pole \\[0.2em]
   $[\gev]$ & $[\gev^2]$ & $[\gev^2]$ &   &   &   &  \\[0.2em]
\hline
2.308  & 2.215 &  0.000 & $2.078 \pm 0.180$ & 2.785 & $1.635 \pm 0.11$ & 0.359 \\[0.2em]
2.264  & 2.279 &  0.037 & $1.365 \pm 0.125$ & 1.692 & $1.395 \pm 0.08$ & 0.243  \\[0.2em]
2.223  & 2.411 &  0.050 & $0.980 \pm 0.110$ & 1.342 & $1.337 \pm 0.08$ & 0.216  \\[0.2em]
2.181  & 2.539 &  0.060 & $0.786 \pm 0.114$ & 1.105 & $1.304 \pm 0.08$ & 0.200 \\[0.2em]
2.127  & 2.703 &  0.087 & $0.564 \pm 0.123$ & 0.775 & $1.240 \pm 0.08$ & 0.164 \\[0.2em]
\hline
\end{tabular}
\end{center}
\caption{The longitudinal and transverse cross sections for $\pi^+$ electroproduction 
at low values of $W$ in $\mu {\rm b}/\gev^2$.  Quoted are the data of \ci{horn06} and the contributions
from the pion pole.}
\label{tab:2}
\renewcommand{\arraystretch}{1.0}   
\end{table*}

\begin{figure}[t]
\begin{center}
\includegraphics[width=0.43\tw,bb=25 348 532 743,clip=true]{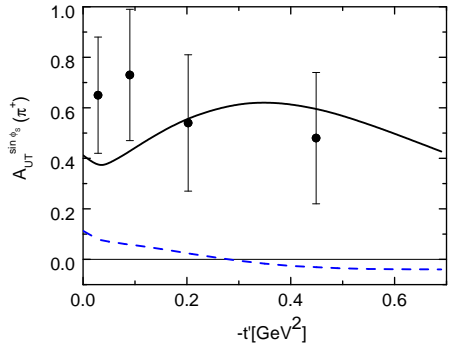}
\caption{\label{fig:electro-aut2-hermes} The $\sin{\phi_s}$ moment for a
  transversely polarized target at $Q^2\simeq 2.45\,\gev^2$ and
  $W=3.99\,\gev$. The prediction from our handbag approach is shown as a solid
  line. The dashed line is obtained disregarding the twist-3 contribution. 
  Data are taken from \ci{Hristova}. (colors online)}
\end{center}
\end{figure}

An even more striking evidence for contributions from transversely polarized
photons comes from the recent HERMES measurement of the asymmetry, $A_{UT}$, 
for a transversely polarized target \ci{Hristova}. Various $\sin{\Phi}$-moments 
can be extracted from the measured differential cross section, where $\Phi$ 
is a linear combination of the azimuthal angle $\phi$ between the lepton and
the hadron plane and the angle $\phi_S$ that describes the orientation of the
target spin vector with respect to the lepton plane. The associated 
coefficients $A_{UT}^{\sin{\Phi}}$ are defined in the appendix; their
relations to the helicity amplitudes are given in \req{eq:AUTs} (cf.\ also
Tab.\ \ref{tab:1}). The $\sin{\phi_S}$ moment is particularly large, only 
mildly $t$ dependent and does not show any indication of a turnover towards 
zero for $t^\prime\to 0$, see Fig.\ \ref{fig:electro-aut2-hermes}. Inspection 
of \req{eq:AUTs} reveals that this behavior of $A_{UT}^{\sin{\phi_S}}$ at
small $-t^\prime$, can only be produced by the longitudinal-transverse 
interference term ${\rm Im}\Big[{\cal M}^*_{0-,++}{\cal M}_{0+,0+}\Big]$. Both 
the contributing amplitudes are helicity non-flip ones and are therefore not 
forced to vanish in the forward direction by angular momentum conservation.
The other interference term occurring for $A_{UT}^{\sin{\phi_S}}$ falls off
proportional to $t^\prime$ for $t^\prime\to 0$ by the same conversation
law. The longitudinal amplitude ${\cal M}_{0+,0+}$ contributing to 
$A_{UT}^{\sin{\phi_S}}$ comprises the asymptotically dominant leading-twist
contribution \req{eq:L-amplitudes}. The other amplitude in this 
interference term, ${\cal M}_{0-,++}$, which as we explained in the 
preceding section, plays a special role for the pion-pole contribution, 
remains to be discussed. It cannot solely be fed by the pole contribution 
\req{eq:pol-cut} since, in this case, it would only interfere with the GPD 
contribution to ${\cal M}_{0+,0+}$ given in \req{eq:L-amplitudes}. As can 
readily  be checked this contribution is too small (see Fig.\ 
\ref{fig:electro-aut2-hermes}). Hence, there must be 
another dynamical mechanism contributing to ${\cal M}_{0-,++}$. 

It is tempting to model the amplitude ${\cal M}_{0-,++}$ from the ordinary
GPDs $\widetilde{H}$, and so on. It is easy to check however that in this 
case the leading-twist pion spin \wf{} $q^\prime\cdot\gamma\,\gamma_5/\sqrt{2}$ 
leads to ${\cal M}_{0-,++}=0$. A more complicated spin \wf{} ivolving quark 
transverse momenta, would lead to a non-vanishing contribution but this one
would  vanish $\propto t^\prime$ for $t^\prime\to 0$. This behavior is a
consequence of the fact the partons emitted and re-absorbed by the nucleon have
the same helicity for these GPDs. Therefore, the hadronic matrix
element for the transition $p(+) q(+)\to n(-) q(+)$ has to disappear as 
$\sqrt{-t^\prime}$ as consequence of angular momentum conservation \ci{diehl01}. 
Moreover, the subprocess is described by the amplitude ${\cal H}_{0+,++}$ in
this case which drops as $\sqrt{-t^\prime}$, too. We conclude: In order to obtain a
non-vanishing $A_{UT}^{\sin{\phi_s}}$ the amplitude ${\cal M}_{0-,++}$  has to
be fed by dynamics that differs from that one taken into account by a
handbag approach which includes only the helicity non-flip GPDs.
This conclusion is conform with the observation that the pion-pole
contribution to the transverse cross section is clearly below experiment \ci{horn06}. 

As a side remark we note that one observes in electroproduction of
pions the same phenomenon as in photoproduction: The
pion pole including the modification \req{eq:pol-cut} is insufficient to 
describe the differential cross section at and near $t=0$. Additional 
contributions to ${\cal M}_{0-,++}$ are required which, in the Regge approach,
are typically modelled as Regge cuts \ci{storrow,worden}. A similar observation can also be 
made in $p\bar{p}\to n \bar{n}$ \ci{leader}. The peculiar dynamics of 
photoproduction is qualitatively transferred to the transverse amplitudes of 
electroproduction in the Regge model. Quantitatively, however, modifications 
may occur since Regge residues and perhaps even the trajectories depend on 
$Q^2$ in general \ci{collins,donnachie}.

%%%%%%%%%%%%%%%%%%%%%%%%%%%%%%%%%%%%%%%%%%%%%%%%%%%%%%%%%%%%%%%%%%%%%%%%%%%%%%
\section{A twist-3 contribution}
\label{sec:twist-3}
%%%%%%%%%%%%%%%%%%%%%%%%%%%%%%%%%%%%%%%%%%%%%%%%%%%%%%%%%%%%%%%%%%%%%%%%%%%%%
There is a second set of four GPDs \ci{diehl01,hoodbhoy} parameterizing the
soft proton matrix elements. The partons emitted and absorbed by the nucleons 
have opposite helicities for these GPDs. Not much is known about them which 
are denoted by $H_T, \widetilde{H}_T, E_T$ and $\widetilde{E}_T$. There are 
only a few papers to be found in the literature in which they have been
applied, e.g.\ in \ci{Ivanov:2002jj} to electroproduction of two vector 
mesons, in \ci{huang04} to wide-angle photoproduction of pions 
or in \ci{liuti08} to $\pi^0$ electroproduction. In contrast to the ordinary 
GPDs which are $\gamma^+(\gamma_5)$ proton matrix elements of quark field 
operators the helicity flip ones are $\sigma^{+j}$ matrix elements where 
$j(=1,2)$ is a transverse index. As one can readily convince oneself by simply 
counting the number of gamma matrices in the relevant lowest-order Feynman 
graphs for the subprocess (see Fig.\ \ref{fig:graphs}), the leading-twist 
pion spin \wf{} $q^\prime\cdot\gamma\,\gamma_5/\sqrt{2}$ leads to a vanishing 
amplitude. Hence, one has to apply the pion's twist-3 spin \wf s.

The contributions of the twist-3 mechanism to the helicity amplitudes read
\ba
{\cal M}^{{\rm twist-3}}_{0+,\mu +} &=& 
                       e_0\frac{\sqrt{-t^\prime}}{2m} \int_{-1}^1 d\xb\,
          \left\{\big({\cal H}_{0+,\mu -}-{\cal H}_{0-,\mu +}\big)\widetilde{H}^{(3)}_T
              \right. \nn\\
          &+& \left. \big[(1-\xi){\cal H}_{0+,\mu -}-(1+\xi){\cal H}_{0-,\mu
              +}\big]E^{(3)}_T/2 \right.\nn\\
         &+& \left. \big[(1-\xi){\cal H}_{0+,\mu -}+(1+\xi){\cal H}_{0-,\mu
             +}\big]\widetilde{E}^{(3)}_T/2
                \right\} \nn\\
{\cal M}^{{\rm twist-3}}_{0-,\mu +} &=& e_0 \sqrt{1-\xi^2}\int_{-1}^1 d\xb\,
             \left\{\big({\cal H}_{0+,\mu -}-{\cal H}_{0-,\mu +}\big)
                        \frac{t^\prime}{4m^2}\widetilde{H}^{(3)}_T \right. \nn\\
            &+&\left. {\cal H}_{0-,\mu +} \big[H^{(3)}_T -\frac{\xi}{1-\xi^2}(\xi E^{(3)}_T-
            \widetilde{E}^{(3)}_T)\big]\right\}\,.
\label{twist3-ampl}
\ea
The amplitudes are generalizations of the wide-angle results derived in
\ci{huang04}. Given that we are interested in rather small $\xi$ and small
$-t^\prime$ the dominant twist-3 contribution comes from $H_T$ and occurs in
the amplitude ${\cal M}_{0-,+ +}$ which importance we have discussed in 
Sec.\ \ref{sec:transversal}. The amplitude ${\cal M}_{0-,- +}$ although also  
fed by $H_T$ is suppressed by the factor $t^\prime/Q^2$ generated by the double
helicity-flip subprocess amplitude ${\cal H}_{0-,-+}$. Also the twist-3
contributions to the longitudinal amplitudes can be neglected; they come
together with the subprocess amplitude ${\cal H}_{0\mp,0\pm}$  and are
therefore suppressed by $\sqrt{-t^\prime}/Q$ as compared to the $H_T$ 
contribution to ${\cal M}_{0-,+ +}$.  

A complete twist-3 pion wave function would include a pseudoscalar and a tensor 
two-particle term and in addition a three-particle contribution \ci{braun90}.
Although it cannot be justified theoretically except for $Q^2\to\infty$, the
latter is usually neglected in applications \ci{huang04,ben00,ball}. We 
also do so in order to accomplish a first, admittedly rough estimate of the 
twist-3 contribution to $\pi^+$ electroproduction. In general, however, the
three-particle contribution may play an important role in reliable estimates
of higher-twist contributions as well as in achieving gauge invariant
amplitudes in general higher-twist scenarios \ci{lech09}.

According to Beneke and Feldmann \ci{ben00}, the light-cone projection
operator of an outgoing pion in momentum space, including the twist-2 and 3 
two-particle contributions, reads 
\ba
{\cal P}_{\alpha \beta,k l}^{\pi(ab)} &=& \frac{f_\pi}{2\sqrt{2 N_c}} \;
{\cal C}_\pi^{a b} \, \frac{\delta_{k l}}{\sqrt{N_c}}
\left\{\frac{\gamma_5}{\sqrt{2}} \, q\sla{}^\prime
  \Phi_\pi(\tau) \, + \, \mu_\pi \, \frac{\gamma_5}{\sqrt{2}}
        \Big[ \Phi_{\pi\,p}(\tau)
                                                 \right.\nn\\
    &&\hspace{-0.0cm} \left. 
           - i\sigma_{\mu\nu} \frac{q^{\prime\,\mu} k^{\prime\,\nu}}{q^\prime\cdot k^\prime} 
              \frac{\Phi^\prime_{\pi\,\sigma}(\tau)}{6}
                    + i \sigma_{\mu\nu} q^{\prime\,\mu}
            \frac{\Phi_{\pi\,\sigma}(\tau)}{6} \frac{
     \partial}{\partial k_{\perp\nu}}\Big]\right\}_{\alpha \beta}\,,  
\label{pion-proj}
\ea
where $\alpha$ ($a$, $k$) and $\beta$ ($b$, $l$) represent Dirac (flavor,
color) labels of the quark and antiquark, respectively. This version of the
projector which is the one proposed in \ci{ben00} conveniently rewritten in 
the notation of \ci{passek}, can be found in \ci{huang04}.
The parameter $\mu_\pi$ is not just the pion mass but it is proportional to
the chiral condensate
\be
\mu_\pi \= m^2_\pi/(m_u+m_d)\,.
\label{eq:condensate}
\ee
The masses $m_u$ and $m_d$ in \req{eq:condensate} are current quark masses \ci{PDG}. At
a scale of $2\,\gev$ $\mu_\pi$ has the familiar value of $\simeq 2\,\gev$. In 
\req{pion-proj}, $\vk$ denotes the transverse momentum of the quark entering 
the meson, defined with respect to the meson's momentum, $q^\prime$. After 
performing the derivative the collinear limit, $\vk=0$, is taken. 
Note that in the massless limit we are working in, the two vectors $q'$ and
$k^\prime$ being the momentum of the outgoing parton,  are light-like
and their space components have opposite sign. The projector takes into
account the familiar twist-2 distribution amplitude $\Phi_\pi(\tau)$ and the 
two-particle twist-3 ones $\Phi_{\pi\,p}(\tau)$ and $\Phi_{\pi\,\sigma}(\tau)$. 
In \req{pion-proj}, $\Phi^\prime_{\pi\,\sigma}$ denotes the derivative of 
$\Phi_{\pi\,\sigma}$ with respect to $\tau$. The use of the twist-3
part of the projector within the modified perturbative approach is fully 
analogous to that of the twist-2 piece, see \ci{GK1,GK2}.  

Assuming the three-particle distributions to be strictly zero, the
equation of motion fix the twist-3 distribution 
amplitudes \ci{braun90, ben00} to:
\be 
\Phi_{\pi\,p}(\tau) \;=\;1\,, \qquad  
\Phi_{\pi\,\sigma}(\tau) \;=\; 6\,\tau\,(1-\tau)\,.
\label{da-special}
\ee

The flavor weight factors ${\cal C}_\pi^{ab}$ comprise the flavor structure
of the meson. They read
\be
{\cal C}_{\pi^0}^{uu}=-\,{\cal C}_{\pi^0}^{dd}=1/\sqrt{2}\,,\qquad 
{\cal C}_{\pi^+}^{ud}={\cal C}_{\pi^-}^{du}=1\,,
\label{eq:factors}
\ee
for charged and uncharged pions. All other weight factors are zero 
(e.g., ${\cal C}^{ss}_{\pi_0}=0$) since the projection operator
\req{pion-proj} implies the valence quark approximation for the meson.

The twist-3 subprocess amplitude ${\cal H}_{0-,++}$ to lowest order of
perturbative QCD is to be calculated from the same set of Feynman graphs as
the twist-2 contribution (see Fig.\ \ref{fig:graphs}). It turns out that the
tensor terms provide contributions which are proportional to $t^\prime/Q^2$  
and are consequently neglected by us. Thus, only the pseudoscalar contribution 
to ${\cal H}_{0-,++}$  remains at small $-t$. As for the twist-2 contributions 
we calculate it within the modified perturbative approach and obtain 
($\bar{\tau}=1-\tau$)
\ba
{\cal H}_{0-,++}^{\pi^+}&=& \frac{\sqrt{2}}{\pi} \frac{C_F}{\sqrt{N_c}} \mu_\pi
                    \int d\tau d^2\vk \Psi_{\pi\,p}(\tau,\vk) \als(\mu_R)\nn\\
               &\times& \left(\frac{e_u}{\xb-\xi+\imath\varepsilon}\,
              \frac1{\bar{\tau}(\xb-\xi)Q^2/(2\xi)-\vk^2+\imath\varepsilon} \right.\nn\\
              &+& \left. \frac{e_d}{\xb+\xi-\imath\varepsilon}\,
              \frac1{\tau(\xb+\xi)Q^2/(2\xi)+\vk^2-\imath\varepsilon}\right)\,.
\label{eq:twist-3-sub}
\ea
In the spirit of the modified perturbative approach we only retain $\vk$ in
the denominators of the parton propagators where it plays a crucial role. Its
square competes with terms $\propto \tau (\bar{\tau}) Q^2$ which become small
in the end-point regions where either $\tau$ or $\bar{\tau}$ tends to zero. 
The expression \req{eq:twist-3-sub} is to be Fourier transformed to the impact
parameter space and there to be multiplied by the Sudakov factor ${\rm exp}[-S]$ 
analogously to \req{mod-amp}. The LCWF corresponding to the pseudoscalar term 
is denoted by $\Psi_{\pi\,p}$. Evidently the twist-3 contribution is 
suppressed by $\mu_\pi/Q$ as compared to the twist-2 ones. The large value of 
$\mu_\pi$ provides the justification for making allowance for this twist-3
effect in our analysis while other contributions being parametrically
suppressed by $\sqrt{-t^\prime}/Q$, are omitted. We stress that, to the order
of accuracy we are working, the amplitude \req{eq:twist-3-sub} respects
electromagnetic and color gauge invariance.

In collinear approximation ${\cal H}_{0-,++}$ is infrared singular since the
corresponding distribution amplitude \req{da-special} does not vanish at the
end points. Moreover, in collinear approximation there is a double pole 
$(\xb-\xi)^2$. It is elucidating to compare with wide-angle photoproduction 
in collinear approximation \ci{huang04,huang} where the Mandelstam  variables 
$t$ and $u$ provide the large scale required by the handbag factorization. In 
collinear approximation the twist-3 contribution is regular in the wide-angle 
region, in fact zero by an exact cancellation of the pseudoscalar and the 
tensor terms, if the three-particle contribution is neglected \ci{huang04}. 
The origin of this difference between deeply-virtual and wide-angle pion 
production lies in propagators of the type
\be
B \propto \frac{t}{\tau t -\bar{\tau}Q^2}\,,
\ee
occurring for the two tensor terms. Hence, in the wide-angle region one has
\be
B \propto 1/\tau\,,
\ee
i.e.\ a contribution that has the same $t$ dependence as the pseudoscalar term
and an exact cancellation may happen and indeed takes place. On the other
hand, for the case of electroproduction, one has 
\be 
B \propto \frac{t}{\bar{\tau}Q^2}\,,
\ee
which cannot cancel the pseudoscalar term in the forward limit.

%%%%%%%%%%%%%%%%%%%%%%%%%%%%%%%%%%%%%%%%%%%%%%%%%%%%%%%%%%%%%%%%%%%%%%%%%%%%%%%%%%%
\section{GPDs and meson wave functions}
\label{sec:GPD}
%%%%%%%%%%%%%%%%%%%%%%%%%%%%%%%%%%%%%%%%%%%%%%%%%%%%%%%%%%%%%%%%%%%%%%%%%%%%%%%%%%%
According to the discussion presented in Sec.\ \ref{sec:leading} we need to
model the GPDs $\widetilde{H}$ and $\widetilde{E}$ for valence quarks. For the
twist-3 contribution we need the GPD $H_T$ in addition. Throughout 
flavor-symmetric sea GPDs are assumed. We also have to discuss the pion LCWFs.

The GPDs $\widetilde{H}$, $\widetilde{E}$ and $H_T$ for valence quarks are
constructed from their double distribution representation. For the latter the
familiar ansatz \ci{mus99}
\be
f_i^a (\rho, \eta,t) \= \exp{[(b_i-\alpha^\prime_i\ln{\rho})t]}\, 
        F_i^a(\rho,\xi=t=0)\,
      \frac{3}{4}\frac{[(1-\rho)^2-\eta^2]}{(1-\rho)^3}\,\Theta{(\rho)} 
\ee
is made which consists of the forward limit, $F_i$, of the relevant GPD, a
weight function and a $t$ dependent factor parameterized in a Regge-like 
fashion with a Regge trajectory, $\alpha(t)$ and a residue function. The $t=0$
part of this factor is absorbed into the forward limit of the GPD. 

As in our previous work \ci{GK3} the forward limit of $\widetilde{H}$, the
polarized parton distributions,  is taken from the analysis presented in
\ci{BB}. The parameters in the $t$ dependent factor are poorly determined 
as yet since the only sensitive experimental information in vector-meson 
electroproduction is provided by the helicity correlation $A_{LL}$ for 
$\rho^0$ electroproduction measured by HERMES \ci{hermes03} and COMPASS 
\ci{compass07}. These data suffer from very large errors. Since $\pi^+$ 
electroproduction offers a better handle on these parameters we fix them 
in the present analysis: $\tilde{\alpha}_h^\prime\=0.45 \gev^{-2}t$ and 
$\tilde{b}_h=0$\,.
  
Nothing is known about $\widetilde{E}$ as yet. In order to simplify matters
we first assume that
\be
\widetilde{E}^u\=-\widetilde{E}^d \= \frac12\widetilde{E}^{(3)}
\ee
holds as for the perturbative part of the pion-pole term \req{pole}. This relation
between the $u$ and $d$ quark GPDs is also similar to the situation for the
GPD $E$ \ci{DFJK4,GK4}. The forward limit of $\widetilde{E}^u$ is
taken as
\be
\tilde{e}^{u}(\rho)\= \widetilde{N}_e  \rho^{-0.48} (1-\rho)^5\,.
\label{E-double-distribution}
\ee
For the $t$ dependence we take the parameters 
$\widetilde{\alpha}_e^\prime=0.25\,\gev^{-2}$ and $\tilde{b}_e=0$. The value
of the normalization $\tilde{N}_e$ is obtained from fits to the $\pi^+$ 
electroproduction data. We obtain $\tilde{N}_e=-25$.

The  contribution from  $\widetilde{E}$ has the opposite sign to the pion-pole
contribution and becomes larger than the latter one for large $-t$.
Our model GPD $\widetilde{E}^{(3)}$ is similar in size and sign to the 
results obtained in \ci{penttinen} where $\widetilde{E}$ has been
calculated within the chiral soliton model. We have also checked that 
$\widetilde{E}$ and $\widetilde{H}$ respect the positivity bounds presented in
\ci{pobylitsa,diehl03} (cf.\ the analogues of (193) and (194) in \ci{diehl03}).  

\begin{figure}[t]
\begin{center}
\includegraphics[width=0.45\tw,bb=91 406 512 749,clip=true]{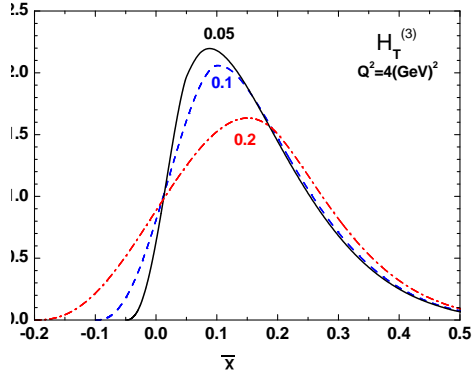}
\caption{\label{fig:HT} The isovector combination of the GPD $H_T$ for
  versus $\xb$ for $\xi=0.05$ (solid), $0.1$ (dashed) and $0.2$
  (dash-dotted line). (colors online)} 
\end{center}
\end{figure}

In order to model the GPD $H_T$ we rely on a model for the transversity distributions
invented in Ref.\ \ci{anselmino} in order to fit the data on the azimuthal
asymmetry in semi-inclusive deep inelastic scattering and in inclusive
two-hadron production in electro-positron annihilation. The model proposed in
\ci{anselmino} assumes that the transversity distributions which are the
forward limit of the GPD $H_T$, are given by 
\be
\delta^q(\rho) \= 7.46\, N^q_{T}\,\rho (1-\rho)^5\, \Big[q(\rho) + \Delta q(\rho)\Big]\,,
\label{HT-double-distribution}
\ee
where $q$ and $\Delta q$ are the usual unpolarized and polarized parton distributions.
In concord with \ci{anselmino} we take for the normalization constants the
values
\be
N^u_{T}\= 0.5\,, \qquad N^d_{T}\=-0.6\,. 
\ee
For the slope of the Regge trajectory we again use $\alpha^\prime_T =0.45\,\gev^{-2}$.
The slope parameter $b_T$ is adjusted to experiment leading to $b_T=0.95\,\gev^{-2}$\,.
The resultant GPD is shown if Fig.\ \ref{fig:HT} for sample values of the skewness. 

For the pion LCWF we use a simple Gaussian \ci{jakob} 
\be
\Psi_\pi\=\frac{\sqrt{6}}{f_\pi} \exp{[-a^2_\pi \vk^2/(\tau\bar{\tau})]}\,,
\label{pi-wf}
\ee
which has been probed in many applications.
The transverse size parameter, $a_\pi$, is fixed from the $\pi^0\to
\gamma\gamma$ decay \ci{brodsky} to be
\be 
a_\pi \= [\sqrt{8} \pi f_\pi]^{-1}\,.
\ee
The LCWF \req{pi-wf} corresponds to the twist-2 asymptotic meson
distribution amplitude $\Phi_{AS}=6\tau\bar{\tau}$. One may consider more
general $\tau$ dependencies by multiplying $\Phi_{AS}$ with a sum of
Gegenbauer terms \ci{brodsky}. Exploiting values of the lowest Gegenbauer
coefficients derived from QCD sum rules \ci{ball,bakulev} or extracted from 
data on the $\pi\gamma$ transition form factor \ci{DKV1} lead to results for 
$\pi^+$ electroproduction which do not differ much from those obtained with
$\Phi_{AS}$. We therefore show only results for the latter case in the
following.
 
For the pseudoscalar LCWF required for the calculation of the twist-3
contribution we take
\be
\Psi_{\pi\,p} \= \frac{16\pi^{3/2}}{\sqrt{6}}\,f_\pi a_\pi^3 k_\perp \exp{[-a_p^2 \vk^2]}\,,
\ee
which corresponds to the distribution amplitude $\Phi_{\pi\,p}$ quoted in
\req{da-special}. The combination of the two requirements, namely the
constant distribution amplitudes and normalizibility of the \wf{} forces the
use of the simple exponential which is to be contrasted with \req{pi-wf}. The
r.m.s.\ value of $\vk$ is just $1/a_p$. In order to have $\langle
\vk^2\rangle^{1/2}\simeq 0.5\,\gev$ a value of $2.0\,\gev^{-1}$ is required for
$a_p$ \ci{jakob}. 

\begin{table*}[t]
\renewcommand{\arraystretch}{1.4} 
\begin{center}
\begin{tabular}{|c|| c | c | c | c |}
\hline     
 observable  & dominant &   amplitudes  & low $t^\prime$ \\
        & interf. term  &   &  behavior \\[0.2em]   
\hline
$A_{UT}^{\sin(\phi-\phi_s)}$ &  LL  & ${\rm Im}\big[{\cal M}^*_{0-,0+}
                       {\cal M}_{0+,0+}\big]$ & $\propto \sqrt{-t^\prime}$   \\[0.2em]
$A_{UT}^{\sin(\phi_s)}$ & LT  &  ${\rm Im}\big[{\cal M}^*_{0-,++}{\cal M}_{0+,0+}\big]$  & const.  \\[0.2em]
$A_{UT}^{\sin(2\phi-\phi_s)}$ & LT & ${\rm Im}\big[{\cal M}^*_{0-,-+}
                             {\cal M}_{0+,0+}\big]^{\;1)}$ &  $\propto t^\prime$\\[0.2em]
$A_{UT}^{\sin(\phi+\phi_s)}$ & TT &  ${\rm Im}\big[{\cal M}^*_{0-,++}
                        {\cal M}_{0+,++}\big]$ & $\propto \sqrt{-t^\prime}$    \\[0.2em]
$A_{UT}^{\sin(2\phi+\phi_s)}$ & TT &  $\propto \sin{\theta_\gamma}$ &  $\propto t^\prime$\\[0.2em]
$A_{UT}^{\sin(3\phi-\phi_s)}$ & TT & ${\rm Im}\big[{\cal M}^*_{0-,-+}
                             {\cal M}_{0+,-+}\big]$ & $\propto (-t^\prime)^{(3/2)}$  \\[0.2em]
\hline
$A_{UL}^{\sin(\phi)}$   & LT & ${\rm Im}\big[{\cal M}^*_{0-,++} {\cal M}_{0-,0+}\big]$ & 
$\propto \sqrt{-t^\prime}$   \\[0.2em]
\hline
\end{tabular}
\end{center}
\caption{Features of the asymmetries for a transversally and longitudinally
  polarized target. The photon polarization is denoted by L (longitudinal) and
  T (transversal). 1) There is a second contribution for which the
  helicities of the outgoing proton are interchanged.}
\label{tab:1}
\renewcommand{\arraystretch}{1.0}   
\end{table*} 

%%%%%%%%%%%%%%%%%%%%%%%%%%%%%%%%%%%%%%%%%%%%%%%%%%%%%%%%%%%%%%%%%%%%%%%%%%%%%%%%%%%
\section{Results from the handbag approach}
\label{sec:fits}
%%%%%%%%%%%%%%%%%%%%%%%%%%%%%%%%%%%%%%%%%%%%%%%%%%%%%%%%%%%%%%%%%%%%%%%%%%%%%%%%%%%

In the preceding sections we have specified all the ingredients of our
handbag approach to $\pi^+$ electroproduction and turn now to the calculation
of observables and the comparison with experiment. The relations between
observables and helicity amplitudes are compiled in the appendix.

It is to be stressed that the calculation of $\pi^+$ electroproduction is
intriguing. Three, as yet practically unknown GPDs are needed. The $t$
dependence of the data \ci{horn06,HERMES07,Hristova,hermes02} is much more
complicated than that of diffractive vector-meson electroproduction
\ci{GK2,GK3}. Is is also important to realize at this point that we model only
one of the transverse amplitudes in detail, namely ${\cal M}_{0-,++}$. For the
other transverse amplitudes we only consider the tiny pion-pole contribution. 
Moreover, the twist-3 effect can only be regarded as a rough estimated since
the three-particle contribution to the pion state is ignored. 
For these reasons, we can only aim at an understanding of the gross feature of
the data. A precise fit of all details of the data is beyond feasibility at
present.

\begin{figure}[t]
\begin{center}
\includegraphics[width=0.45\tw,bb=39 350 532 743,clip=true]{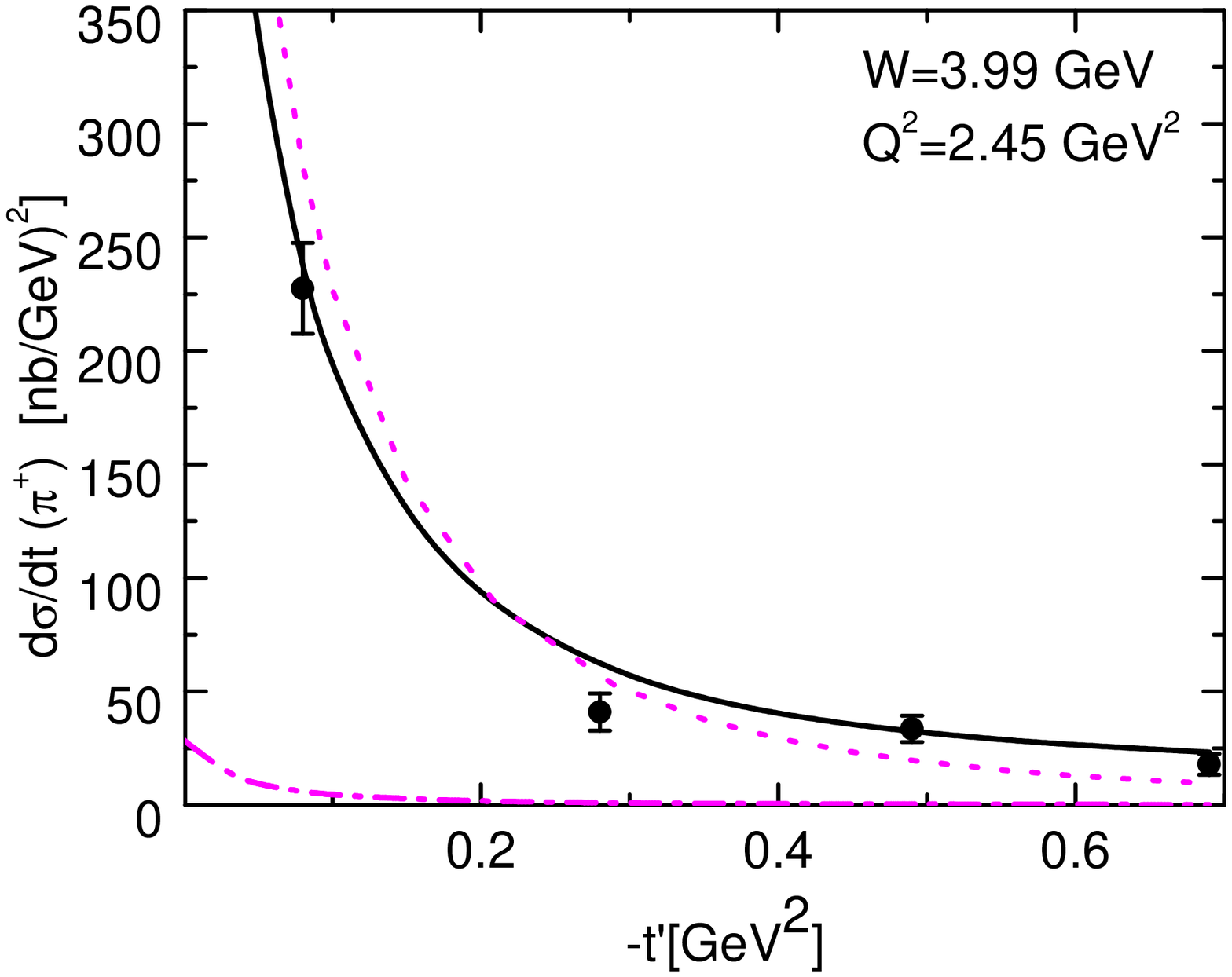}
\includegraphics[width=0.45\tw,bb=42 349 531 744,clip=true]{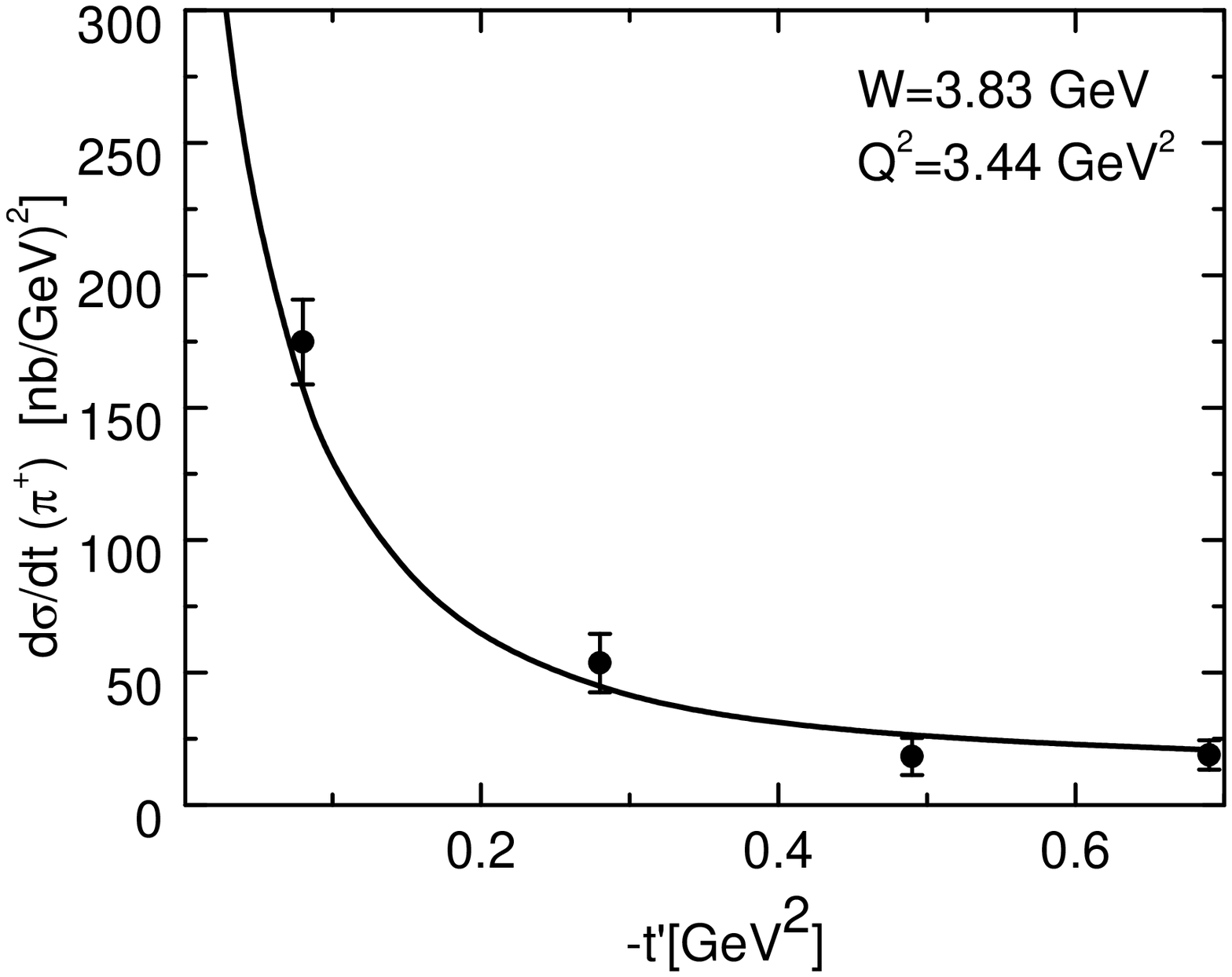}
\caption{\label{fig:electro-cross-hermes-1} The un-separated $\pi^+$-electroproduction
  cross section versus $-t^\prime$. The solid lines represent our
  predictions for the un-separated cross section. The dashed and
  dot-dashed lines are the pion-pole contributions for the un-separated
  and transversal cross sections, respectively. Data are taken from
  \ci{HERMES07}. (colors online)}  
\end{center}
\end{figure}

\begin{figure}[t]
\begin{center}
\includegraphics[width=0.45\tw,bb=36 350 531 742,clip=true]{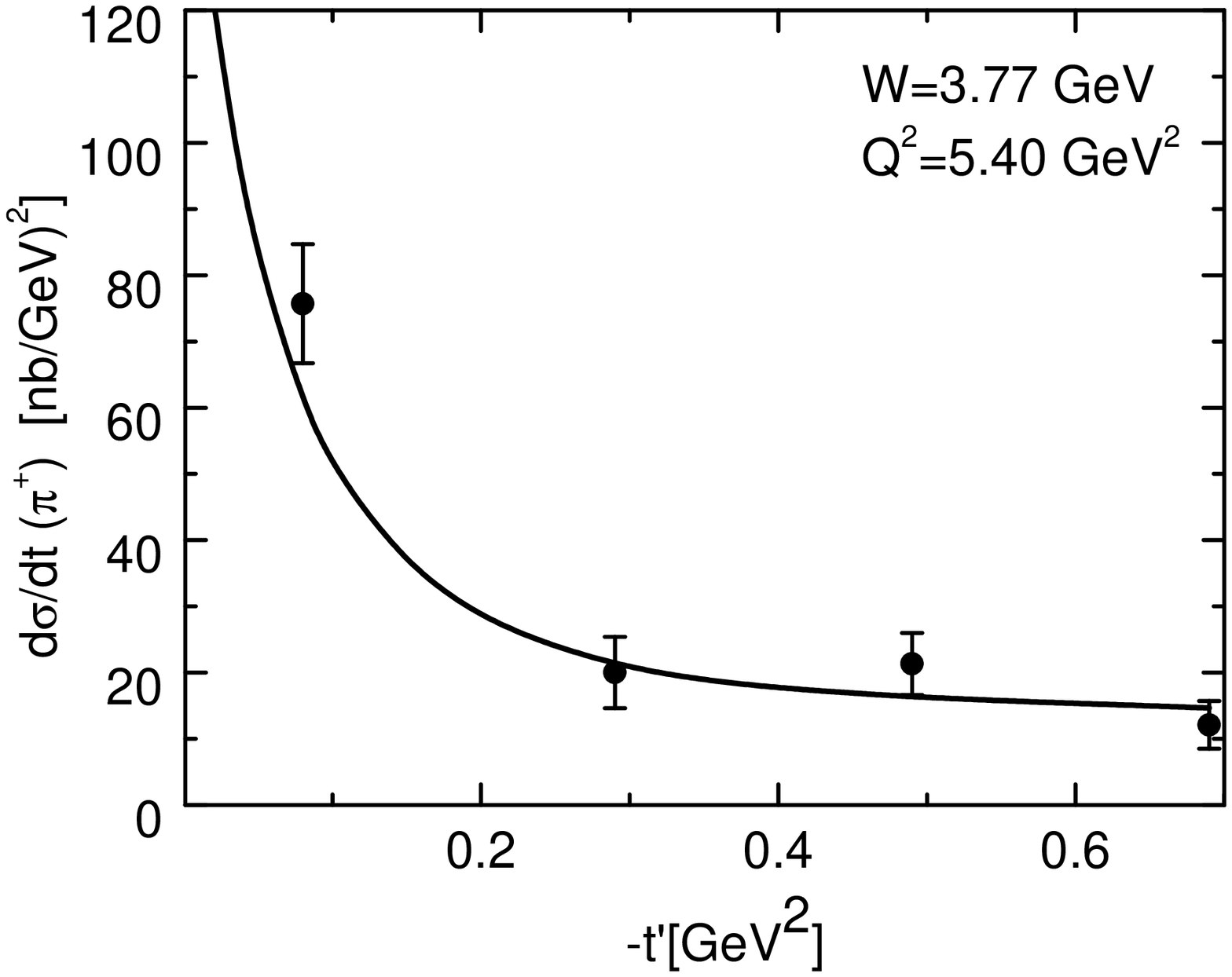}
\includegraphics[width=0.455\tw,bb=32 361 533 752,clip=true]{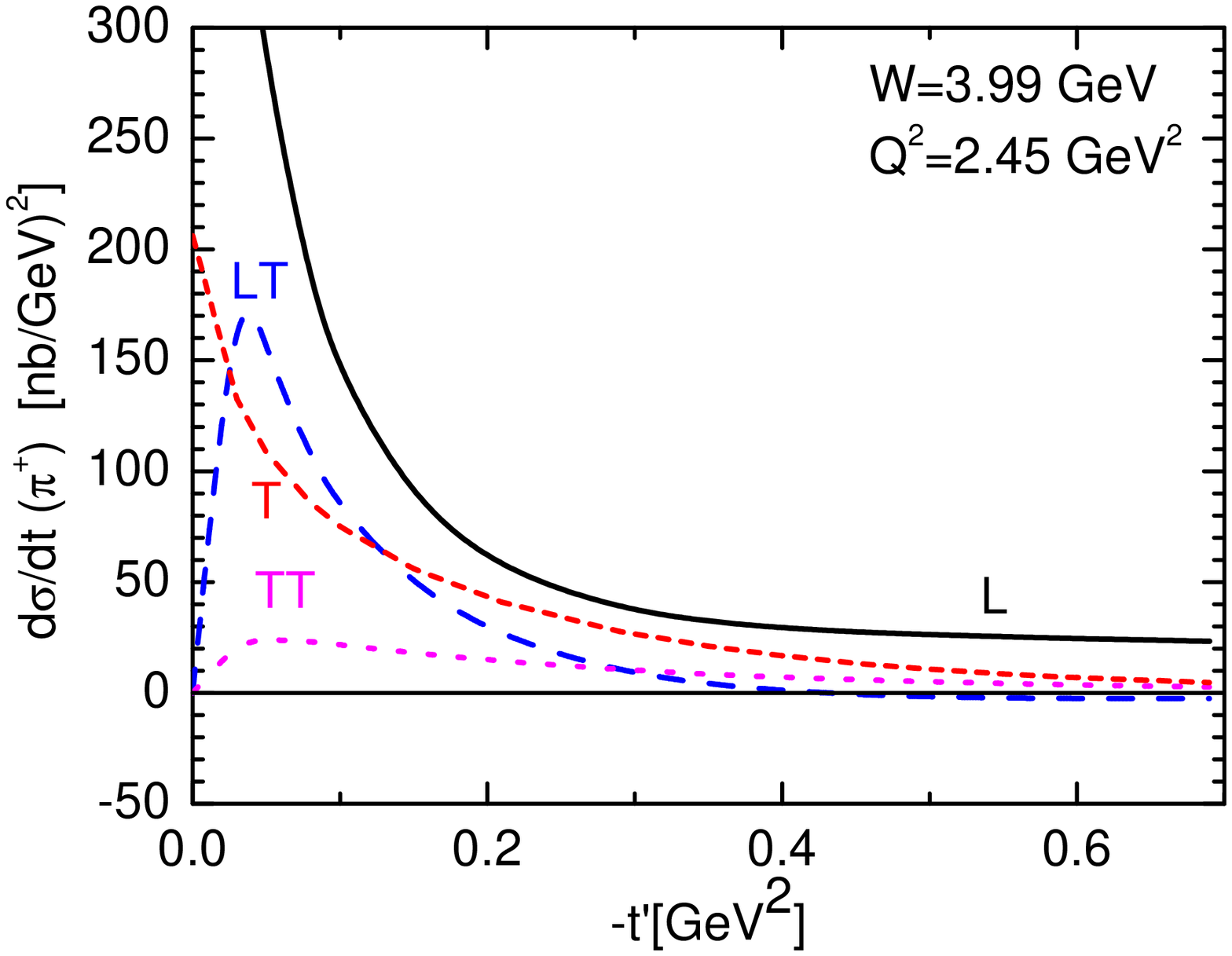}
%\vspace*{-0.45\tw}
\caption{Left: As Fig.\ \ref{fig:electro-cross-hermes-1} but for $Q^2=5.4\,\gev^2$ and
  $W=3.77\,\gev$. Right: The partial cross sections $d\sigma_L/dt$ (solid line),
  $d\sigma_T/dt$ (short dashed line), $d\sigma_{LT}/dt$ (long dashed
  line) and $d\sigma_{TT}/dt$ (dotted line) at $Q^2=2.45\,\gev^2$ and
  $W=3.99\,\gev$. (colors  online)}  
\label{fig:electro-cross-hermes-2}
\end{center}
\end{figure}

In Figs.\ \ref{fig:electro-cross-hermes-1} and \ref{fig:electro-cross-hermes-2}  
we show the un-separated differential cross section $d\sigma_T/dt + \eps
d\sigma_L/dt$ at various values of $Q^2$ and $W$. We take $\eps=0.8$ for the
polarization of the virtual photon, a value that is characteristic of HERMES
kinematics. As the comparison with the HERMES \ci{HERMES07} data reveals we
achieve a fair description of the cross section. 

In Fig.\ \ref{fig:electro-cross-hermes-2} we also display the four partial cross
sections. The longitudinal cross section is large and drops down rapidly with
increasing $-t^\prime$. Also the transverse cross section, essentially made up
by the twist-3 mechanism is rather large. Hence, a considerable share
of the un-separated cross section measured by HERMES \ci{HERMES07} is
due to contributions from transversely polarized photons. 
The longitudinal-transverse interference $d\sigma_{LT}/dt$ is particularly
large at very small $-t^\prime$  due to the interference between the pion-pole
contribution to ${\cal M}_{0-,0+}$ and the twist-3 contribution to 
${\cal M}_{0-,++}$. Since the first amplitude vanishes for forward scattering 
a pronounced bump is produced by the interference term ${\rm Re}\big[{\cal
  M}^*_{0-,++} {\cal M}_{0-,0+}\big]$. For large  $-t^\prime$ the $LT$ cross section 
becomes negative. Not unexpectedly the TT cross section is very small.      

\begin{figure}[t]
\begin{center}
\includegraphics[width=0.40\tw,bb=16 349 533 743,clip=true]{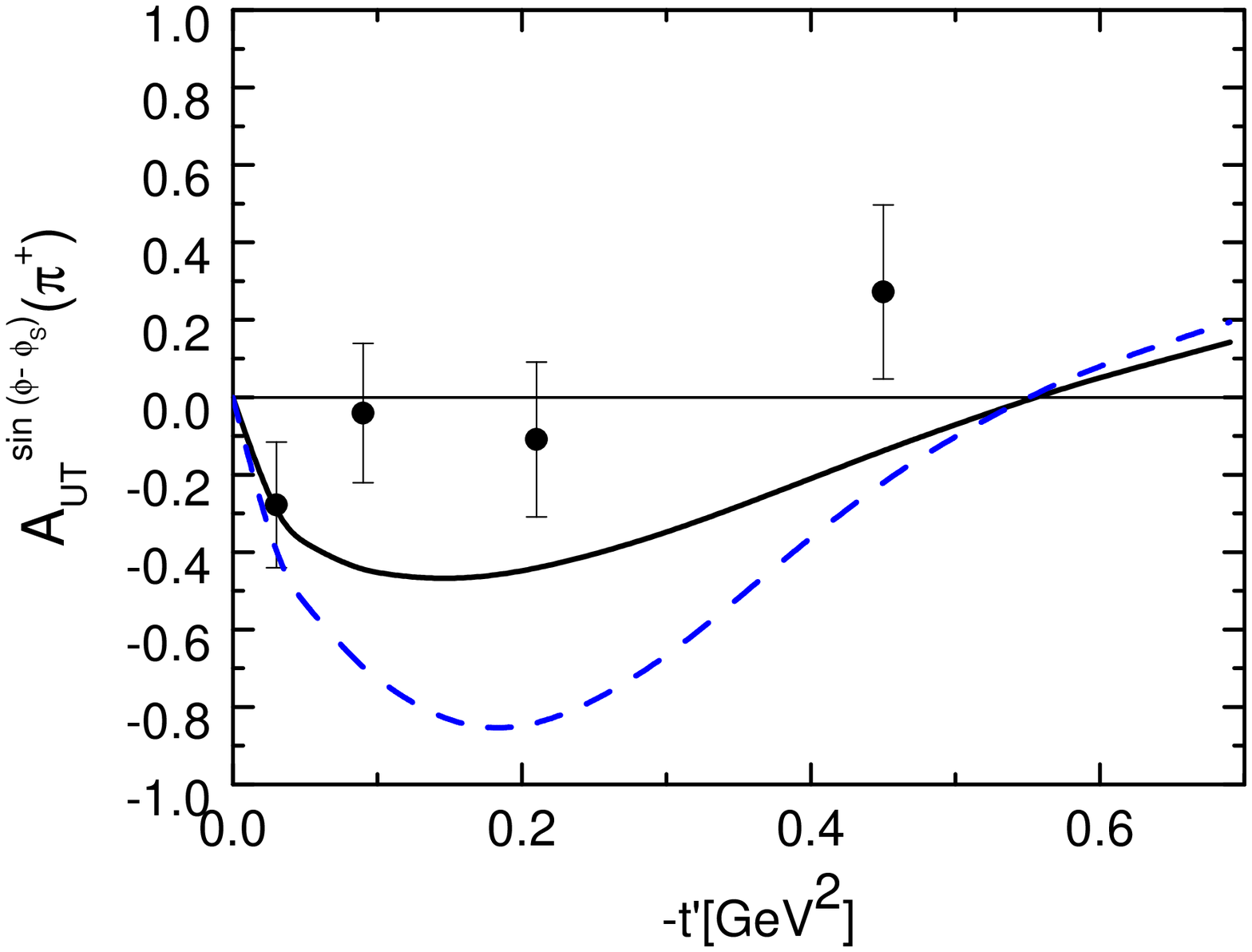}
\includegraphics[width=0.40\tw,bb=20 349 533 743,clip=true]{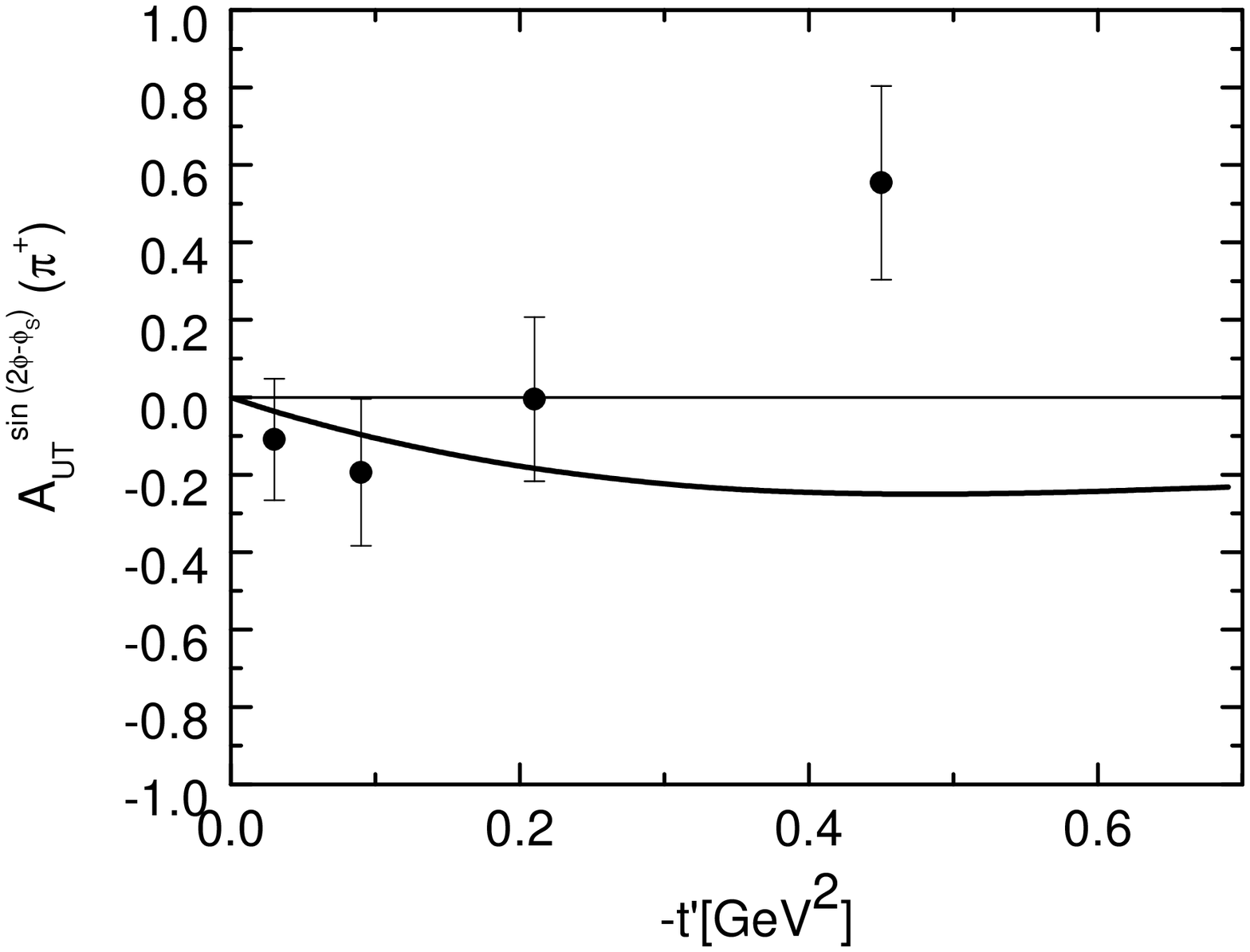}
\caption{Predictions for the  $\sin{(\phi-\phi_s)}$ (left) and the
  $\sin{(2\phi-\phi_s)}$ (right) moment at $Q^2=2.45\,\gev^2$ and
  $W=3.99\,\gev$ shown as solid lines. The dashed line represents the
  longitudinal contribution to the $\sin{(\phi-\phi_s)}$ moment. 
  Data are taken from \ci{Hristova}. (colors online)}  
\label{fig:electro-aut-div-hermes}
\end{center}
\end{figure}

Results for the asymmetries $A_{UT}$ obtained with a transversely
polarized proton target, are displayed in Figs.\ \ref{fig:electro-aut2-hermes} 
and \ref{fig:electro-aut-div-hermes}. In order to elucidate the behavior of
the target asymmetries it is advisable to simplify the expressions
\req{eq:AUTs} and \req{eq:sin_phi} and to inspect the most prominent
contributions. They and their behavior for $t^\prime\to 0$ are listed in 
Tab.\ \ref{tab:1}. The three moments $\sin{(\phi+\phi_s)}$, 
$\sin{(2\phi+\phi_s)}$ and $\sin{(3\phi-\phi_s)}$ 
are only fed by transverse-transverse interference terms and are therefore
small in our approach. This result is in concord with the smallness of the
$TT$ cross section. The hierarchy of the moments at fixed $Q^2$ and $W$   
perceptible in Tab.\ \ref{tab:1} is in agreement with the findings of the
HERMES collaboration \ci{Hristova}. Particularly large is the $\sin{\phi_s}$
moment, cf.\ Fig.\ \ref{fig:electro-aut2-hermes}. This is a consequence of the
fact that it is determined by the interference of two large helicity non-flip
amplitudes as we discussed in Sect.\ \ref{sec:transversal}. The change of sign
of the $\sin{(\phi-\phi_s)}$ moment at large $-t^\prime$ (see Fig.\ 
\ref{fig:electro-aut-div-hermes}) is due to the contribution from $\widetilde{E}$ 
which over-compensates the pole contribution to ${\cal M}_{0-,0+}$
there. In order to reproduce this feature of the $A_{UT}$ data a large
contribution from $\widetilde{E}$ is demanded. In absolute value the
$\sin{(\phi-\phi_s)}$ moment is somewhat large in our approach as compared to
experiment. In Fig.\ \ref{fig:electro-aut-div-hermes} we also show a
result for this moment obtained by neglecting all contributions from
transversely polarized photons. In this case the prediction is much
too large in absolute value. The $\sin{(2\phi-\phi_s)}$ moment is
small as is the experimental value. Only the change of the sign at 
large $-t^\prime $ is not reproduced by
us. This observable however is given by an interference terms between the 
longitudinal amplitudes and transversal one other than ${\cal M}_{0-,++}$. 
Thus, likely an improvement of this moment would require a detailed modeling
of the small transverse amplitudes. In summary, it is fair to conclude that
the pattern of the experimental results on the target asymmetries
\ci{Hristova} is fully in agreement with the theoretical hierarchy. We finally
mention that the $Q^2$ dependence of our predictions for the $A_{UT}$ parameters
at fixed $t^\prime$ and $W$ is very smooth, typically the parameters decrease
by about $10\%$ in absolute value between $Q^2=2.5$ and $6\,\gev^2$. 

\begin{figure}[t]
\begin{center}
\includegraphics[width=0.42\tw,bb=33 351 533 746,clip=true]{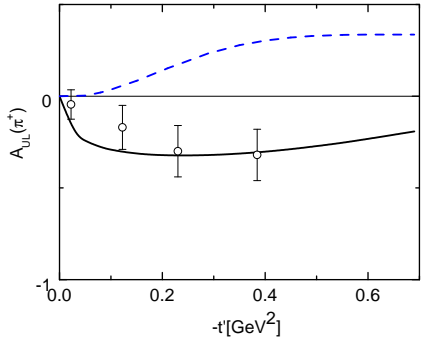}
%\vspace*{-0.45\tw}
\caption{The asymmetry for a longitudinally polarized target at
$Q^2\simeq 2.4\,\gev^2$ and $W\simeq 4.1\,\gev$. The dashed line is 
obtained disregarding the twist-3 contribution. Data are taken from 
\ci{hermes02}. (colors online)}  
\label{fig:electro-aul}
\end{center}
\end{figure}

The HERMES collaboration has also measured the  asymmetry, $A_{UL}$ for a
longitudinally polarized target \ci{hermes02}. It is important to realize that
$A_{UL}$ is dominated by the interference term (see \req{eq:sin_phi} and Tab.\
\ref{tab:1})
\be
A_{UL} \propto {\rm Im}\Big[{\cal M}_{0-,++}^* {\cal M}_{0-,0+}\Big]\,.
\label{eq:inter}
\ee 
The second term $\propto  {\cal M}_{0+,0+}$ in \req{eq:sin_phi} is zero since
the combination $ {\cal M}_{0+,++}+ {\cal M}_{0+,-+}$ is zero in our approach, see
\req{eq:pion-pole-amplitude}. The interference term in \req{eq:inter} is
similar to the one being responsible for the $\sin{(\phi_s)}$ moment obtained
with a transversally polarized target. Only that instead of the longitudinal
helicity non-flip amplitude, the helicty flip one occurs with the consequence 
of a small $t^\prime$ behavior as $\propto \sqrt{-t^\prime}$ for $A_{UL}$
making it smaller than $A_{UL}^{\sin{(\phi_s)}}$ in absolute value.
In Fig.\ \ref{fig:electro-aul} we compare our results with the HERMES data on
$A_{UL}$ \ci{hermes02}. A good agreement between theory and experiment
can be seen. In Figs.\ \ref{fig:electro-aut2-hermes} and
\ref{fig:electro-aul} we also display results which are obtained by
disregarding the twist-3 contribution. The important role of the
twist-3 contribution for the interpretation of the spin asymmetries is
clearly seen.

The asymmetry obtained with a longitudinally polarized beam is dominated by
the interference term \req{eq:inter}, too. But this asymmetry is smaller than
$A_{UL}$ by the factor $\sqrt{(1-\epsilon)/(1+\epsilon)}\simeq 1/3$.

Although the main purpose of this article is focussed on the analysis of the
HERMES data
one may be also interested in comparing our approach with the Jefferson Lab
data on the cross sections \ci{horn06}. However, a word of caution is
advisable. It is known that the cross section for $\rho^0$ production
shows an asymmetric minimum at $W\simeq 4\,\gev$ as a function of
energy but at fixed $Q^2$. Between 2 and $4\,\gev$ the cross section
falls down by nearly an order of magnitude \ci{clas-rho} while above 
$4\,\gev$ it rises slowly, see \ci{GK2} and references therein. On the 
other hand, the cross section for $\phi$ production increases for all
energies above $2\,\gev$ \ci{GK2,clas-phi,HERMES-prel}. The handbag
approach proposed by us \ci{GK2,GK3} is in perfect agreement with the
slow increase of the cross sections for $\rho^0$ and $\phi$ production
but the sharp rise below $4\,\gev$ is not reproduced. What
dynamical mechanism is responsible for that feature of the data is unknown as
yet. One may surmise that the valence quarks are responsible for this
peculiar behavior at low $W$ since this effect is not seen in the
cross section for $\phi$ production. Since $\pi^+$ production is solely 
fed by the valence quarks one may expect a similar failure of our
handbag approach as for $\rho^0$ production. In order to examine whether or
not this is the case we work out the longitudinal and transverse cross
sections within our approach at the kinematics of the $F_\pi-2$
experiment. We find reasonable agreement for the transverse cross section 
with the Jefferson Lab data \ci{horn06} while the longitudinal cross section
is somewhat too small as compared to experiment. This failure is caused by the
strong increase of the contribution from $\widetilde{E}$ with increasing 
skewness in comparison to the pion-pole contribution due to the extra factor 
of $\xi$ in \req{eq:L-amplitudes} which leads to a strong cancellation of 
these two terms. A strong $\widetilde{E}$ is required by the HERMES data on
the spin asymmetries \ci{Hristova,hermes02}. Thus, we see that the application
of our handbag approach at small $W$, i.e. large $\xi$, is problematic 
although not as drastic as for the case of the $\rho^0$. It however
remains to be answered whether this difficulty is due to our particular 
model GPDs or whether one sees here other dynamics which cannot be
incorporated into the handbag approach like in the case of the $\rho^0$. 
We stress that our approach is
designed for small skewness. At larger values of it our
parameterization of the GPDs \req{E-double-distribution} and
\req{HT-double-distribution} are perhaps to simple and may require
improvements. Also the neglect of the helicity-flip GPDs $E_T$ and
$\widetilde{E}_T$ may not be justified at large skewness. Given all
these uncertainties of our handbag approach we think that the results
for low $W$ are promising although a detailed investigation of this
kinematical region is still necessary. 

%%%%%%%%%%%%%%%%%%%%%%%%%%%%%%%%%%%%%%%%%%%%%%%%%%%%%%%%%%%%%%%%%%%%%%%%%%%%
\section{Electroproduction of $\pi^0$ mesons}
\label{sec:pi0}
%%%%%%%%%%%%%%%%%%%%%%%%%%%%%%%%%%%%%%%%%%%%%%%%%%%%%%%%%%%%%%%%%%%%%%%%%%%
Here, in this section, we are going to comment briefly on hard electroproduction of
uncharged pions in the light of our findings for $\pi^+$ production. Now,
instead of the isovector combination, we need $e_u F^u-e_d F^d$ where $F$ is
one of the GPDs $\widetilde{H}$, $\widetilde{E}$ or $H_T$. Since all the
involved GPDs have opposite signs for $u$ and $d$ valence quarks there is a
partial cancellation of the two contribution in contrast to $\pi^+$
production. Also the pion pole does not contribute in this case. Hence, a much
smaller cross section is expected for $\pi^0$ production than for the case of
$\pi^+$. On the other hand, a detailed experimental and theoretical study of 
$\pi^0$ electroproduction may provide distinct information on the relevant GPDs 
in the absence of the strong contributions from the pion pole.

The amplitudes now read (the flavor weight factors are defined in \req{eq:factors})
\ba
{\cal M}_{0+,0+}^{\pi^0} &=& \sqrt{1-\xi^2}\, \frac{e_0}{Q}\,\sum_{a=u,d} e_a\,
{\cal C}_{\pi^0}^{aa} \Big[
                    \;\langle \widetilde{H}^{a} \rangle
            -\frac{\xi^2}{1-\xi^2} \;\langle \widetilde{E}^{a} \rangle \Big]\,,\nn\\
{\cal M}_{0-,0+}^{\pi^0} &=&
\frac{\sqrt{t_0-t}}{2m}\,\xi  \frac{e_0}{Q}  \;\sum_{a=u,d}\,e_a\, {\cal C}_{\pi^0}^{aa}
\langle\widetilde{E}^{a} \rangle\,, \nn\\
{\cal M}_{0-,++}^{\pi^0} &=& e_0\sqrt{1-\xi^2}\,\frac{\mu_\pi}{Q^2}\,\sum_{a=u,d}
   e_a \,{\cal C}_{\pi^0}^{aa}\; \langle H_T^{a} \rangle\,.
\label{eq:pi0-amplitudes}
\ea
With the exception of the amplitudes which are related to those given in
\req{eq:pi0-amplitudes} by parity conservation, all other amplitudes are zero. 
Only valence quarks contribute electroproduction of uncharged pions.

The convolutions in \req{eq:pi0-amplitudes} are given by expressions analogous to 
\req{eq:convolution} with subprocess amplitudes like \req{mod-amp} or 
\req{eq:twist-3-sub}. The only difference is that the quark fractional charges
$e_a$ now appear in combination with the GPDs and not in the subprocess amplitudes 
(for more details cf.\ \ci{GK4}). In the spirit of the modified perturbative 
approach we have to Fourier transform this expression and to multiply it with 
the Sudakov factor in the impact parameter space as in \req{mod-amp}. 

\begin{figure}[ht]
\begin{center}
\includegraphics[width=0.45\tw,bb=50 350 533 743,clip=true]{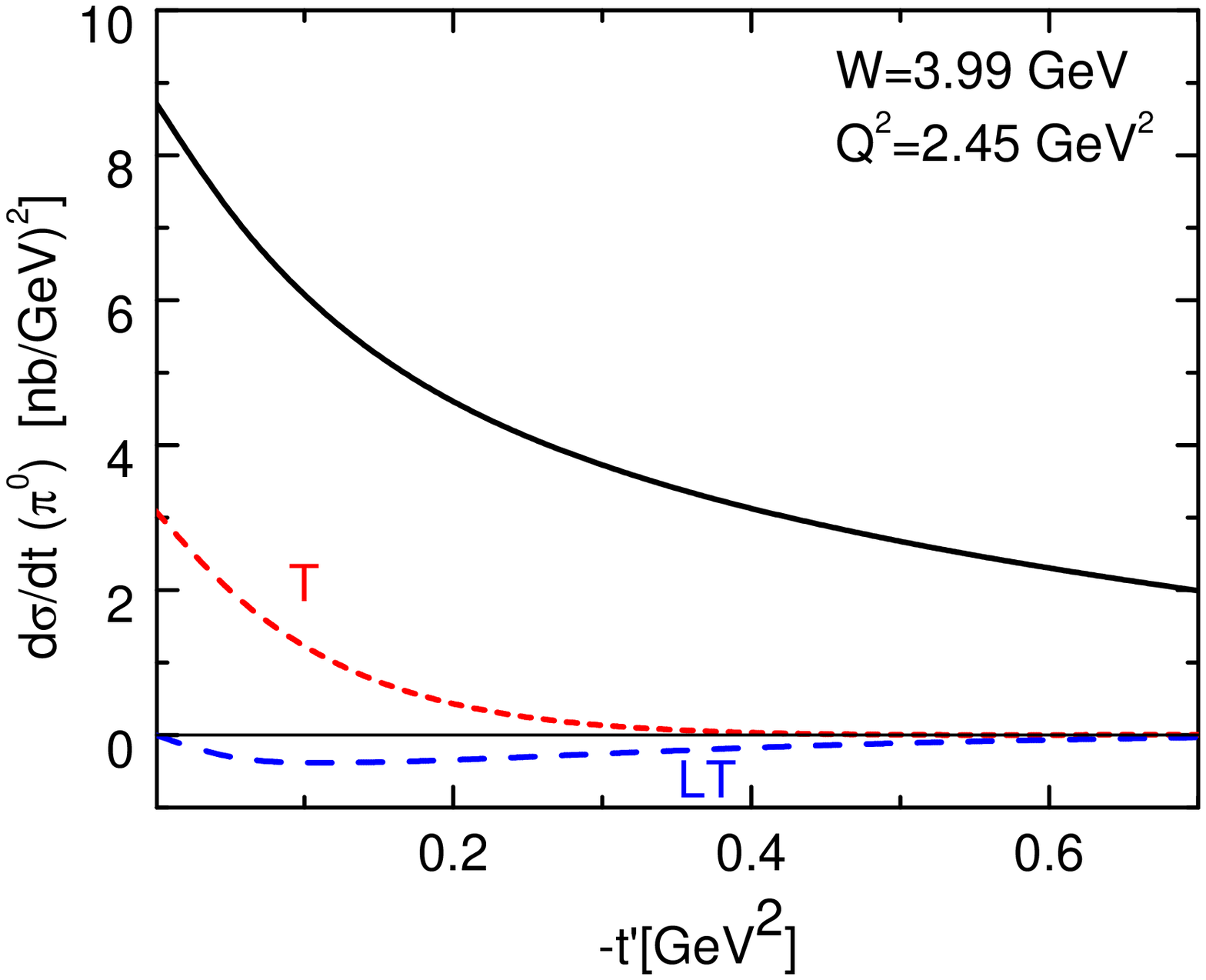}
\includegraphics[width=0.46\tw,bb=31 350 533 743,clip=true]{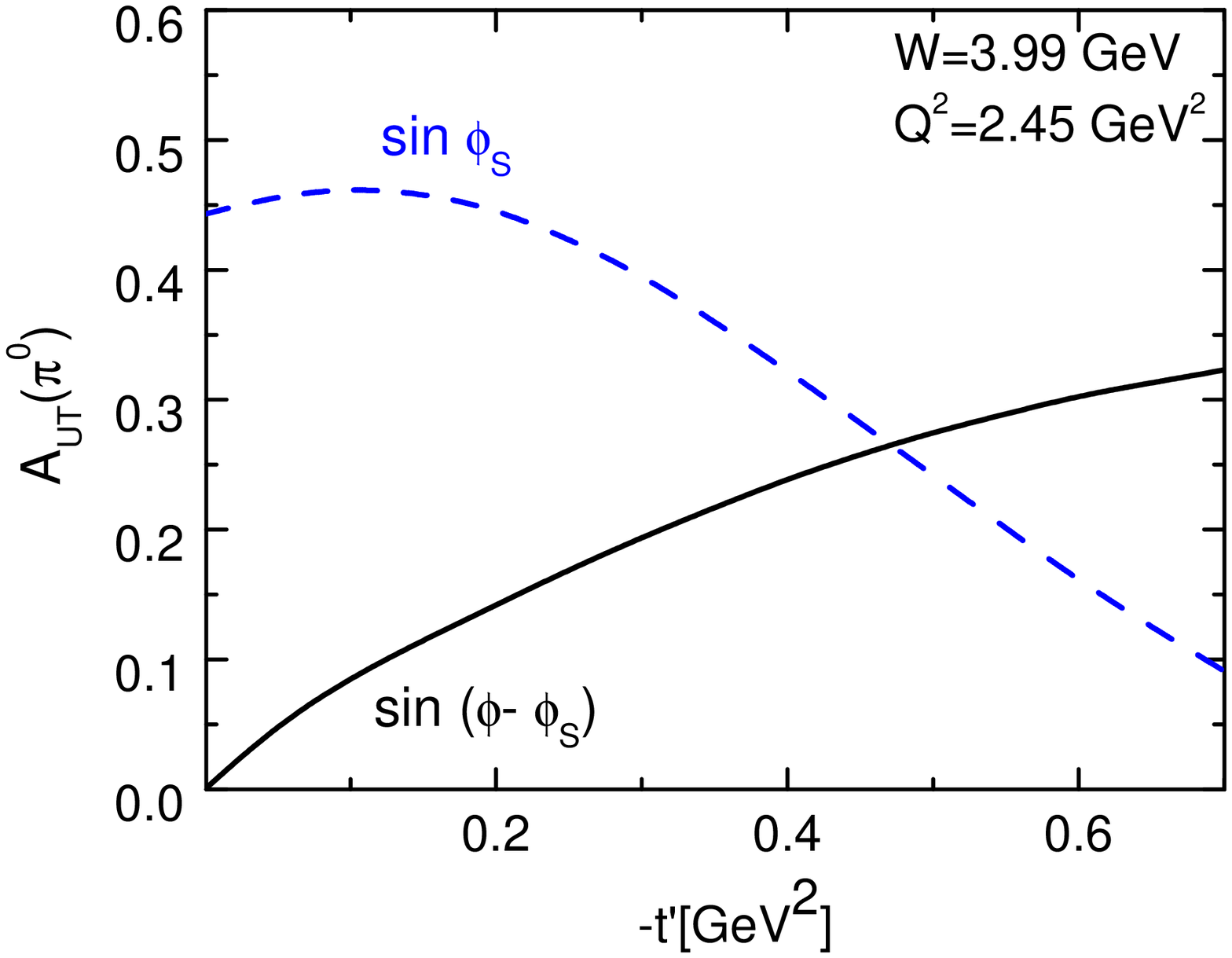}
\caption{Predictions for the cross section (left) and $A_{UT}$ (right) for  
  $\pi^0$ electroproduction versus $-t^\prime$ at  $W=3.99\,\gev$ and 
  $Q^2=2.45\,\gev^2$. The un-separated (transverse, longitudinal-transverse 
  interference) cross section is shown as a solid (dashed, dotted) line. The 
  ${\sin{(\phi-\phi_s)}}$ moment is shown as a solid line, the $\sin{\phi_s}$ 
  moment as a dashed one. (colors online)} 
\label{fig:pi0-cross-hermes}
\end{center}
\end{figure}

In Fig.\ \ref{fig:pi0-cross-hermes} we present a few predictions for the
un-separated cross section and the $A_{UT}^{\sin{(\phi-\phi_s)}}$ asymmetry for
$\pi^0$ electroproduction. For typical HERMES kinematics the longitudinal 
cross section is dominant. We also see that the LT interference cross section
is very small and exhibits a smooth $t$ dependence. The pronounced peak at
small $-t^\prime$ seen for $\pi^+$ production (see Fig.\
\ref{fig:electro-cross-hermes-2}) is absent. The $\sin{(\phi-\phi_s)}$ moment has
the opposite sign as for $\pi^+$ production. The reason for these differences
is the absence of the pion pole in $\pi^0$ electroproduction. For small
energies, for instance at $W\simeq 2.2\,\gev$, but the same value of $Q^2$
the cross section for $\pi^0$ production is more than a factor of 10 larger
than at $W\simeq 4\,\gev$. At this low energy the transverse cross section
strongly dominates. 

Our treatment of the twist-3 contribution differs from the approach proposed 
in \ci{liuti08} markedly. In the latter work the subprocess is viewed as
form factors for photon-pion transitions under the action of vector and
axial-vector currents. This is to be contrasted with our perturbative
calculation which leads to a different helicity structure of the twist-3
contribution. Despite this strong contributions from transverse photons to 
electroproduction of pions are also found in \ci{liuti08} at $W\simeq
2.2\,\gev$ and $Q^2\simeq 2.4\,\gev^2$. 

%%%%%%%%%%%%%%%%%%%%%%%%%%%%%%%%%%%%%%%%%%%%%%%%%%%%%%%%%%%%%%%%%%%%%%%%%%%%%%
\section{Twist-3 effects in vector-meson electroproduction}
\label{sec:vector}
%%%%%%%%%%%%%%%%%%%%%%%%%%%%%%%%%%%%%%%%%%%%%%%%%%%%%%%%%%%%%%%%%%%%%%%%%%%%%
The twist-3 mechanism may also be applied to the amplitude ${\cal  M}_{0-,++}$ for
vector-meson production where the helicity label 0 now refers to a longitudinally
polarized vector meson. This amplitude contributes to some of the spin density
matrix elements (SDME), e.g.\ $r_{00}^5$ in the notation of Schilling and Wolf
\ci{schilling}, and influences also the transverse cross section. It is to be
stressed that the parameter $\mu_\pi$ is now to be replaced by the
vector-meson mass, $m_V$. The twist-3 effect is therefore parametrically of 
order $m_V/Q$ which is not much larger than the generally neglected 
$\sqrt{-t^\prime}/Q$ effects.  The results presented in this section are
therefore to be taken with a grain of salt, they are rather of qualitative nature.

The calculation of this twist-3 contribution to hard exclusive vector-meson 
electroproduction is fully analogous to the case of the pion. The
subprocess amplitude is now to be calculated with the covariant spin wave 
function~\footnote{
We remind the reader that the polarization vector of a helicity-zero 
vector meson is $\eps(0)=q^\prime/m_V$ up to corrections of order $m_V/Q$.}
$m_V/\sqrt{2}$ the analogue of the relevant twist-3 piece for pions (see
\req{pion-proj}). It turns out that the subprocess amplitude ${\cal H}_{0-,++}$ is
identically to \req{eq:twist-3-sub} with the obvious replacements of $\mu_\pi$
by $m_V$ and the pseudoscalar pion \wf{} by the corresponding vector-meson one,
$\Psi_{VP}$. Thus, the twist-3 contribution from the valence quarks to
$\rho^0$ and $\omega$ production in momentum space reads
\be
{\cal M}^{V}_{0-,++} \= e_0 \sqrt{1-\xi^2} \,\frac{m_V}{Q^2}\; 
        \sum_{a=u,d} e_a\,{\cal C}^{aa}_V \;\langle H_T^a \rangle 
\label{twist-3-ampl-V}
\ee
in full analogy to \req{eq:pi0-amplitudes}. The flavor weight factors read 
\be
{\cal C}_{\rho^0}^{uu} \= -{\cal C}_{\rho^0}^{dd}\={\cal C}_{\omega}^{uu}\=
{\cal C}_{\omega}^{dd}\=1/\sqrt{2}
\label{eq:flavor}
\ee
for the vector mesons of interest in this work. 

As we already mentioned the twist-3 effect in vector-meson electroproduction
is smaller than in pion production since the large parameter $\mu_\pi$ is to
be replaced by the vector-meson mass. Moreover, there is a partial
cancellation between the $u$ and $d$ valence quark contributions since,
according to \req{eq:flavor}, both partake in the combination  
$e_u H^u_T - e_d H^d_T$ as for the case of the $\pi^0$. Numerical evaluation 
indeed reveals that the twist-3 
mechanism has no perceptible effect in $\rho^0$ electroproduction.  
For $\omega$ production the combination $e_u H^u_T + e_d H^d_T$ occurs which 
entails a larger twist-3 contribution than for the case of the $\rho^0$. 
Still the twist-3 effects is hardly noticeable. As an example we quote the ratio 
of the longitudinal and transverse cross sections for $\omega$ production. At 
$W=5\,\gev$ and $Q^2=3\,\gev^2$ the ratio amounts to 1.52 without and 1.4 with
the twist-3 contribution. Thus, we can conclude that the results on
vector-meson electroproduction presented in \ci{GK2,GK3} remain essentially
unchanged by the inclusion of our twist-3 contribution. A detailed
investigation of twist-3 and perhaps twist-4 contributions to vector-meson
electroproduction is perhaps in order. Such an analysis which is beyond the
scope of the present work, should also include the twist-3 mechanism for the  
gluonic subprocess and transversally polarized vector mesons.

%%%%%%%%%%%%%%%%%%%%%%%%%%%%%%%%%%%%%%%%%%%%%%%%%%%%%%%%%%%%%%%%%%%%%%%%%%%%
\section{Remarks on a symmetry property}
\label{sec:symmetry}
%%%%%%%%%%%%%%%%%%%%%%%%%%%%%%%%%%%%%%%%%%%%%%%%%%%%%%%%%%%%%%%%%%%%%%%%%%%
It is well known that for natural ($N$) and unnatural ($U$) parity exchange
the helicity amplitudes for meson electroproduction satisfy the relations 
\ba
{\cal M}^{MN}_{-\mu'\nu',-\mu\nu} &=& \phantom{-}\eta_M (-1)^{\mu-\mu'}\,{\cal
  M}^{MN}_{\mu'\nu',\mu\nu}\,,\nn\\
{\cal M}^{MU}_{-\mu'\nu',-\mu\nu} &=& -\eta_M(-1)^{\mu-\mu'}\,{\cal
  M}^{MU}_{\mu'\nu',\mu\nu}\,,
\label{eq:natural-unnatural}
 \ea  
where, for vector mesons, $\eta_M$ is +1 while for pseudoscalar mesons it is -1.
In \ci{GK1,GK3} we have shown with the help of parity invariance
\be
{\cal M}^{M}_{-\mu'-\nu',-\mu-\nu} \= \eta_M (-1)^{\mu-\nu -\mu'+\nu'}\,
    {\cal M}^{M}_{\mu'\nu',\mu\nu}\,,
\ee
that at the twist-2 level (with or without power corrections like $\vk$
effects) the handbag amplitudes have the property \req{eq:natural-unnatural}
too. The contributions from the GPDs $H$ and $E$ ($\widetilde{H}$ and 
$\widetilde{E}$) corresponds to (un)natural parity exchange. The absence of 
contributions from $H$ and $E$ in the longitudinal amplitudes for pion 
production \req{eq:L-amplitudes} and \req{eq:pi0-amplitudes} is also a 
consequence of \req{eq:natural-unnatural}. Obviously, the pion-pole amplitudes 
\req{eq:pion-pole-amplitude} corresponds to unnatural parity exchange. On the 
other hand, the modified pole amplitude ${\cal M}^{\rm pole}_{0-,++}$ 
\req{eq:pol-cut} as well as the twist-3 contributions to ${\cal M}_{0-,++}$ 
given in \req{twist3-ampl}, \req{eq:pi0-amplitudes} and \req{twist-3-ampl-V} do
not possess the symmetry property \req{eq:natural-unnatural}. This readily
follows from the fact that the subprocess amplitude ${\cal H}_{0-,-+}$ is 
double-flip one which vanishes at least $\propto t^\prime$ for $t^\prime\to 0$ 
in contrast to the non-flip amplitude ${\cal H}_{0-,++}$. Closer inspection of 
\req{twist3-ampl} reveals that not only contributions from $H_T$ which we 
consider here in this work, do not possess the property \req{eq:natural-unnatural} 
but also those from $E_T$ and $\widetilde{E}_T$ while contributions from 
$\widetilde{H}_T$ behave like natural parity exchange.

%%%%%%%%%%%%%%%%%%%%%%%%%%%%%%%%%%%%%%%%%%%%%%%%%%%%%%%%%%%%%%%%%%%%%%%%%%%%
\section{Summary}
\label{sec:summary}
%%%%%%%%%%%%%%%%%%%%%%%%%%%%%%%%%%%%%%%%%%%%%%%%%%%%%%%%%%%%%%%%%%%%%%%%%%%
Exclusive electroproduction of $\pi^+$ mesons at large $Q^2$ is investigated 
within the handbag approach. It is to be stressed that the process cannot be
understood without taking into account pion exchange properly, i.e.\ the full
experimentally measured electromagnetic form factor of the pion is to be
employed; its asymptotically leading perturbative contribution is
insufficient. In addition to the pion pole and the GPDs $\widetilde{H}$ and 
$\widetilde{E}$ a twist-3 contribution to the amplitude ${\cal M}_{0-,++}$ is
required by the polarization data. In order to estimate the size of this
effect which represents a correction of order $\mu_\pi/Q$, we rely on a 
mechanism that consists of the helicity-flip GPD $H_T$ and the twist-3 pion
wave function. For the latter item only the two-particle partonic states are
considered. This is perhaps a drastic approximation which admittedly allows
only for a rather rough estimate of the size of the twist-3 contribution. All 
the subprocess amplitudes are calculated within the modified perturbative 
approach. Higher order perturbative corrections other than those included in 
the Sudakov factor ( and in the experimental pion form factor) are not taken 
into account. According to Diehl and Kugler \ci{kugler} the first order 
perturbative corrections are rather large for the cross sections for values 
of $Q^2$ around $3\,\gev^2$. However, we remind the reader that the
corrections calculated by Diehl and Kugler refer to the leading-twist 
contribution only, i.e.\ to about $10\%$ of the cross section. Finally
we want to mention that our approach applies to exclusive $\pi^-$
electroproduction as well.\\

%%%%%%%%%%%%%%%%%%%%%%%%%%%%%%%%%%%%%%%%%%%%%%%%%%%%%%%%%%%%%%%%%%%%%%%%%%%%
{\bf Acknowledgements}  We thank M.~Diehl, I.~Hristova and W.-D.~Nowak 
for discussions. We are also grateful to the HERMES collaboration for
providing us their data prior to publication.
This work is supported  in part by the Russian Foundation for
Basic Research, Grant 09-02-01149 and by the Heisenberg-Landau
program and by the BMBF, contract number 06RY258.\\

%%%%%%%%%%%%%%%%%%%%%%%%%%%%%%%%%%%%%%%%%%%%%%%%%%%%%%%%%%%%%%%%%%%%%%%
\begin{appendix}
%%%%%%%%%%%%%%%%%%%%%%%%%%%%%%%%%%%%%%%%%%%%%%%%%%%%%%%%%%%%%%%%%%%%%%
\noindent
{\Large\bf Appendix: Observables for $ep\to e\pi^+n$}\\

The unpolarized $ep\to e\pi^+n$ cross section can be decomposed into a number
of partial cross sections which are observables of the process $\gamma^*p\to\pi^+n$
\ba
\frac{d^4\sigma}{dQ^2dtd\phi} &=& \frac{\ale (s-m^2)}{16\pi^2E_L^2m^2Q^2(1-\eps)}
      \,\left(\frac{d\sigma_T}{dt} +\eps \frac{d\sigma_L}{dt}\right.\nn\\
       &+&\left. \eps\cos{2\phi}\,\frac{d\sigma_{TT}}{dt}
             +\sqrt{2\eps(1+\eps)}\cos{\phi}\frac{d\sigma_{LT}}{dt}\right)\,.
\label{partial-cross-sections}
\ea
The partial cross are expressed in terms of the $\gamma^*p\to\pi^+n$ helicity
amplitudes as follows
\ba
\frac{d\sigma_L}{dt} &=& \frac{\mid {\cal M}_{0+,0+}\mid^2 + \mid{\cal M}_{0-,0+}\mid^2}
                 {16\pi(W^2-m^2)\sqrt{\Lambda(W^2,-Q^2,m^2)}}\,,\nn\\[0.3em]
\frac{d\sigma_T}{dt} &=& \frac{\mid {\cal M}_{0-,++}\mid^2 + \mid{\cal M}_{0-,-+}\mid^2
              +\mid {\cal M}_{0+,++}\mid^2 + \mid{\cal M}_{0+,-+}\mid^2  }
                 {32\pi(W^2-m^2)\sqrt{\Lambda(W^2,-Q^2,m^2)}}\,,\nn\\[0.3em]
\frac{d\sigma_{LT}}{dt} &=& -\frac{\sqrt{2}}           
                 {32\pi(W^2-m^2)\sqrt{\Lambda(W^2,-Q^2,m^2)}}  \nn\\[0.3em]
                           &{\rm Re}&\hspace*{-0.02\tw}\Big[ {\cal M}^*_{0-,0+} 
                           \Big({\cal M}_{0-,++}-{\cal M}_{0-,-+} \Big)
          + {\cal M}^*_{0+,0+} \Big({\cal M}_{0+,++}-{\cal M}_{0+,-+} \Big) \Big]    
                             \,,\nn\\[0.3em]
\frac{d\sigma_{TT}}{dt} &=& -\frac{{\rm Re}\Big[ {\cal M}^*_{0-,++}{\cal M}_{0-,-+}
    + {\cal M}^*_{0+,++}{\cal M}_{0+,-+}  \Big]}
                 {32\pi(W^2-m^2)\sqrt{\Lambda(W^2,-Q^2,m^2)}}\,,
\ea
where $L$ and $T$ label longitudinally and transversally polarized virtual
cross sections. The $ep$ cross section for an unpolarized beam but a
transversally polarized target has been given by Diehl and Sapeta \ci{sapeta}.
It consists of six terms which can be isolated by taking various $\sin$-moments  
\ba
A_{UT}^{\sin(\phi-\phi_s)} \sigma_0&=& -2\epsilon\cos{\theta_\gamma}\,
 {\rm Im} \Big[{\cal M}^*_{0-,0+}{\cal M}_{0+,0+}\Big] \nn\\
&& -\cos{\theta_\gamma}\,{\rm Im}\Big[{\cal M}^*_{0+,++} {\cal M}_{0-,-+}
                            -{\cal M}^*_{0-,++} {\cal
			      M}_{0+,-+}\Big]\,,\nn\\
 &&+\frac12\sin{\theta_\gamma}\sqrt{\epsilon(1+\epsilon)}\,
 {\rm Im}\Big[({\cal M}^*_{0+,++}+ {\cal M}^*_{0+,-+}){\cal M}_{0+,0+} \nn\\
 &&+ ({\cal M}^*_{0-,++}+ {\cal M}^*_{0-,-+}){\cal M}_{0-,0+}\Big]\nn\\[0.3em]
A_{UT}^{\sin(\phi_s)} \sigma_0 &=& \cos{\theta_\gamma}\sqrt{\epsilon(1+\epsilon)}\,
  {\rm Im}\Big[{\cal M}^*_{0+,++}{\cal M}_{0-,0+}
              - {\cal M}^*_{0-,++}{\cal M}_{0+,0+}\Big]\,,\nn\\[0.3em]
A_{UT}^{\sin(2\phi-\phi_s)} \sigma_0 &=& \cos{\theta_\gamma}\sqrt{\epsilon(1+\epsilon)}\,
{\rm Im}\Big[({\cal M}^*_{0+,-+}{\cal M}_{0-,0+}-{\cal M}^*_{0-,-+}{\cal M}_{0+,0+}\Big]\nn\\
&& + \frac12 \epsilon \sin{\theta_\gamma}\, {\rm Im}\Big[{\cal M}^*_{0+,++}{\cal M}_{0+,-+}
                         + {\cal M}^*_{0-,++}{\cal
			   M}_{0-,-+}\Big]\,,\nn\\[0.3em] 
A_{UT}^{\sin(\phi+\phi_s)} \sigma_0  &=& \epsilon\cos{\theta_\gamma}\, 
{\rm Im}\Big[{\cal M}^*_{0-,++} {\cal M}_{0+,++}\Big] \nn\\
&&   +\frac12\sin{\theta_\gamma}\sqrt{\epsilon(1+\epsilon)}\,
 {\rm Im}\Big[({\cal M}^*_{0+,++}+ {\cal M}^*_{0+,-+}){\cal M}_{0+,0+}\nn\\
 &&     + ({\cal M}^*_{0-,++}+ {\cal M}^*_{0-,-+}){\cal M}_{0-,0+}\Big]
\nn\\[0.3em] 
A_{UT}^{\sin(2\phi+\phi_s)} \sigma_0 &=&\frac12\epsilon  \sin{\theta_\gamma}\,
 {\rm Im}\Big[{\cal M}^*_{0+,++}{\cal M}_{0+,-+}+ 
                {\cal M}^*_{0-,++} {\cal M}_{0-,-+}\Big]\,,\nn\\[0.3em]
A_{UT}^{\sin(3\phi-\phi_s)} \sigma_0 &=&\epsilon\cos{\theta_\gamma}\,
   {\rm Im}\Big[{\cal M}^*_{0+,-+}{\cal M}_{0-,-+}\Big]\,.
\label{eq:AUTs}
\ea  
Here, $\phi$ is the azimuthal angle between the lepton and hadron plane and
$\phi_s$ specifies the orientation of the target spin vector with respect to
the lepton plane. The normalization $\sigma_0$ reads
\ba
\sigma_0&=&\frac12 \Big[\mid{\cal M}_{0+,++}\mid^2+\mid{\cal M}_{0-,-+}\mid^2
     +\mid{\cal M}_{0-,++}\mid^2+\mid{\cal M}_{0+,-+}\mid^2\Big] \nn\\
    &+& \eps\; \Big[\mid{\cal M}_{0+,0+}\mid^2+\mid{\cal M}_{0-,0+}\mid^2\Big]\,.
\ea
The angle $\theta_\gamma$ describes the rotation in the lepton plane
from the direction of the incoming lepton to the the virtual photon
one. It is given by \ci{sapeta}
\be
\cos{\theta_\gamma} \=
\sqrt{1-\gamma^2\,\frac{1-y-(y\gamma/2)^2}{1+\gamma^2}}
\simeq 1 -\frac12\gamma^2 (1-y)\,,
\ee
where $\gamma=2\xbj m/Q$ and $y=(W^2+Q^2-m^2)/(s-m^2)$. Finally, $\eps$ is the
ratio of the longitudinal and transversal photon fluxes. 
%For HERMES
%$s\simeq 50 \,\gev^2$ and hence $y\simeq 0.1-0.2$. Thus, one may use 
%$\cos{\theta_\gamma} \simeq 1$.

Another observable of interest is the asymmetry for a longitudinally polarized
target which is obtained from a $\sin{\phi}$-moment od the electroproduction
cross section. In terms of the $\gamma^*p\to\pi^+n$ helicity amplitudes it
reads
\ba
A_{UL} \sigma_0&=&
       - \cos{\theta_\gamma}\,\sqrt{\epsilon(1+\epsilon)}\,
   {\rm Im}\Big[\Big({\cal M}^*_{0+,++}+{\cal M}^*_{0+,-+}\Big){\cal M}_{0+,0+} \nn\\[0.3em]
    &&+ \Big({\cal M}^*_{0-,++}+{\cal M}^*_{0-,-+}\Big){\cal M}_{0-,0+}\Big]\nn\\[0.3em]
    &&+ \sin{\theta_\gamma}\,\Big[2\epsilon\,{\rm Im}\Big({\cal M}^*_{0-,0+}{\cal M}_{0+,0+}\Big)
       -\epsilon\,{\rm Im}\Big({\cal M}^*_{0-,++}{\cal M}_{0+,++}\Big) \nn\\[0.3em]
       &&-{\rm Im}\Big({\cal M}^*_{0-,++}{\cal M}_{0+,-+}-{\cal
	 M}^*_{0+,++}{\cal M}_{0-,-+}\Big)\Big]\,.
\label{eq:sin_phi}
\ea
This leads to a contribution to the lepton-proton cross section as 
\be
  d\sigma \sim P_L \sin{\phi} A_{UL}
\ee
where $P_L$ is the target polarisation. The corresponding longitudinal beam
polarization ($P_l$) reads
\ba
A_{LU} \sigma_0&=&
       - \cos{\theta_\gamma}\,\sqrt{\epsilon(1-\epsilon)}\,
   {\rm Im}\Big[\Big({\cal M}^*_{0+,++}-{\cal M}^*_{0+,-+}\Big){\cal M}_{0+,0+} \nn\\[0.3em]
    &&+ \Big({\cal M}^*_{0-,++}-{\cal M}^*_{0-,-+}\Big){\cal M}_{0-,0+}\Big]\,.
\ea
It contributes to the lepton-proton cross section as
\be
  d\sigma \sim P_l \sin{\phi} A_{LU}\,.
\ee

\end{appendix}
\vskip 10mm 
%%%%%%%%%%%%%%%%%%%%%%%%%%%%%%%%%%%%%%%%%%%%%%%%%%%%%%%%%%%%%%%%%%%%%%

\end{document}